\documentclass[usegraphicx,useAMS,usenatbib]{mn2e}

 \usepackage{times}

\newcommand{\msyv}{M67}
\newcommand{\tuc}{47\,Tuc}
\newcommand{\dss}{$\delta$~Scuti}

\newcommand{\str}{Str\"omgren}
\newcommand{\momf}{{\sc momf}}
\newcommand{\msun}{$M_\odot$}
\newcommand{\teff}{$T_{\rm eff}$}

\newcommand{\logg}{$\log g$}
\newcommand{\feh}{[Fe/H]}

\newcommand{\vsini}{$v \sin i$}
\newcommand{\kms}{km\,s$^{-1}$}
\newcommand{\lee}{{\em left}}
\newcommand{\rii}{{\em right}}
\newcommand{\panel}{{\em panel}}
\newcommand{\panels}{{\em panels}}

\newcommand{\cday}{c\,d$^{-1}$}
\newcommand{\ie}{ie.}
\newcommand{\cf}{cf.}
\newcommand{\eg}{eg.}
\newcommand{\vs}{versus}
\newcommand{\sig}{$\sigma$}
\newcommand{\period}{{\sevensize PERIOD04}}
\newcommand{\mhz}{$\mu$Hz}

\newcommand{\fid}{\sigma_{\rm int, fiduc}}
\newcommand{\isd}{$\sigma_{\rm int}$}
\newcommand{\wscat}{W_{\rm scat}}
\newcommand{\wout}{W_{\rm out}}
\newcommand{\cmd}{colour-magnitude diagram}

\newcommand{\midd}{{\em middle}}
\newcommand{\topp}{{\em top}}
\newcommand{\bott}{{\em bottom}}
\newcommand{\refstar}{$r$}
\newcommand{\refstarth}{$r$th}
\newcommand{\refid}{r}

\newcommand{\apj}{ApJ}
\newcommand{\pasp}{PASP}
\newcommand{\mnras}{MNRAS}
\newcommand{\aap}{A\& A}
\newcommand{\aaps}{A\& AS}

\title[$\delta$ Scuti pulsations in blue stragglers]{Multisite campaign on the open cluster \msyv. 
 III.  $\delta$ Scuti pulsations in the blue stragglers}

\author[H.\ Bruntt et al.]{H.\ Bruntt$^{1}$,  
D.\ Stello$^{1,2,3}$,
J.\ C.\ Su{\'a}rez$^{4,5}$,
T.\ Arentoft$^{2,6}$,
T.\ R.\ Bedding$^{1}$,\newauthor
M.\ Y.\ Bouzid$^{7}$,
Z.\ Csubry$^{8}$,
T.~H.\ Dall$^{9,10}$,
Z.~E.\ Dind$^{1}$,
S.~Frandsen$^{2,6}$,\newauthor
R.\ L.\ Gilliland$^{11}$,
A.\ P.\ Jacob$^{1}$,
H.\ R.\ Jensen$^{2}$,
Y.\ B.\ Kang$^{12}$,
S.-L.\ Kim$^{13}$,\newauthor 
L.\ L.\ Kiss$^{1}$,
H.\ Kjeldsen$^{2,6}$,
J.-R.\ Koo$^{12}$,
J.-A.\ Lee$^{13}$, 
C.-U.\ Lee$^{13}$, 
J.\ Nuspl$^{8}$,\newauthor
C.\ Sterken$^{7}$,
 and 
R.~Szab\'o$^{8,14}$\\
$^{1}$School of Physics, University of Sydney, NSW 2006, Australia\\
$^{2}$Institut for Fysik og Astronomi (IFA), University of Aarhus, DK-8000 Aarhus, Denmark\\
$^{3}$Department of Physics, US Air Force Academy, Colorado Springs, CO 80840, USA\\
$^{4}$Instituto de Astrof\'{\i}sica de Andaluc\'{\i}a, CSIC, CP3004, Granada, Spain\\ 
$^{5}$Observatoire de Paris, LESIA, UMR 8109, Meudon, France\\
$^{6}$Danish AsteroSeismology Centre, University of Aarhus, DK-8000 Aarhus, Denmark\\
$^{7}$Vrije Universiteit Brussel, Pleinlaan 2, B-1050 Brussels, Belgium\\
$^{8}$Konkoly Observatory of the Hungarian Academy of Sciences, H-1525 Budapest, PO Box 67, Hungary\\
$^{9}$Gemini Observatory, 670 N.\ A'ohoku Pl., Hilo, HI 96720, USA\\
$^{10}$European Southern Observatory, Casilla 19001, Santiago 19, Chile\\
$^{11}$Space Telescope Science Institute, 3700 San Martin Dr., Baltimore, USA\\
$^{12}$Department of Astronomy and Space Science, Chungnam National University, Daejeon 305-764, Korea\\
$^{13}$Korea Astronomy and Space Science Institute, Daejeon 305-348, Korea\\
$^{14}$Physics Department, University of Florida, Gainesville, FL, 32611, USA}
\begin{document}

\date{Accepted xxx.yyy 2007. Received hhh.lll 2007}
\pagerange{\pageref{firstpage}--\pageref{lastpage}} \pubyear{2007}
\maketitle
\label{firstpage}

\begin{abstract}
We have made an asteroseismic analysis of the variable blue stragglers in the open cluster \msyv. 
The data set consists of photometric time series from eight sites
using nine 0.6--2.1 meter telescopes with a time baseline of 43 days. 
In two stars, EW~Cnc and EX~Cnc, we detect the 
highest number of frequencies (41 and 26) detected in \dss\ stars
belonging to a stellar cluster, 
and EW~Cnc has the second highest number of frequencies detected in any \dss\ star.
We have computed a grid of pulsation models that take the effects of rotation into account. 
The distribution of observed and theoretical frequencies
show that in a wide frequency range a significant fraction 
of the radial and non-radial low-degree modes are excited to detectable amplitudes.
Despite the large number of observed frequencies we
cannot constrain the fundamental parameters of the stars.
To make progress we need to identify the degrees of some of the modes either
from multi-colour photometry or spectroscopy.
\end{abstract}

\begin{keywords}
stars: individual: EX~Cnc, EW~Cnc,
stars: blue stragglers, 
stars: variables: $\delta$~Scuti, 
open clusters: individual: \msyv\ (NGC 2682)
\end{keywords}

\section{Introduction\label{sec:intro}}

We present the asteroseismic analysis of a new data set comprising 
photometry of the old open cluster \msyv\ (NGC~2682) 
from nine telescopes during a 43-day campaign 
with a total of 100 telescope nights (Stello et al.\ 2006; hereafter Paper~I).  
The main goal of the campaign was to detect oscillations in the
stars on the giant branch (Stello et al.\ 2007; hereafter Paper~II).                 
Here we analyse the variability of the blue straggler (BS) population.

BS stars are defined as being
bluer and more luminous than the turn-off stars in their parent cluster, and 
they are found in all open and globular clusters where dedicated searches have been made.
BSs are important objects since they are directly linked to the interaction between
stellar evolution of binaries and the cluster dynamics.
In the cores of globular clusters the stellar density is high and direct stellar 
collisions may explain the existence of some of the BS stars \citep{davies95}.
However, in open clusters the density of stars is much lower 
and stellar collisions are rare \citep{mardling01}.
BSs in open clusters are therefore generally thought to be formed 
as the gradual coalescence of binary stars \citep{tian06}. 
\cite{hurley05} made a simulation
of \msyv\ that took into account stellar and binary star evolution, in addition 
to the dyna\-mical evolution of the cluster. 
Their simulation  showed that around half of the BSs were formed
from primordial binary systems and the other half from binary stars that had been 
perturbed by close encounters with other stars. 

\begin{figure}
\includegraphics[width=8.8cm]{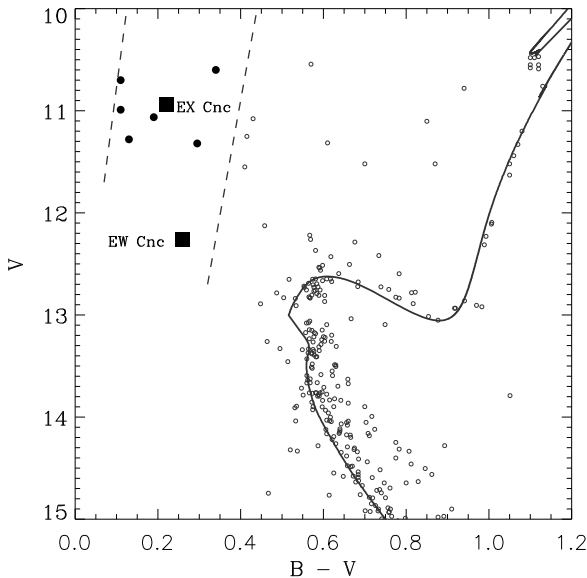}
\caption{The colour-magnitude diagram of \msyv\ based on
standard $B-V$ and $V$ magnitudes.
The solid line is an isochrone for an age of 4~Gyr.
The locations of EW~Cnc and EX~Cnc have been marked with boxes
and six other blue stragglers in the instability strip (dashed lines) 
are marked with filled circles.
\label{fig:cmd}}
\end{figure}

While several of the known BSs in clusters are located near 
the main sequence inside the classical instability strip, 
only around $30$\% show \dss\ pulsations \citep{gill98}.
These objects are particularly interesting 
because their masses can in principle be inferred from comparison with
pulsation models. \cite{bruntt01} studied the variable BSs
in the globular cluster \tuc, 
and could determine the mass of one of them to within $3.5$\% 
from a comparison with models in the Petersen diagram.

We have made a search for oscillations in eight BS stars in
the instability strip in \msyv\ (Fig.~\ref{fig:cmd}).
In particular, we have studied the two known variable BS stars
EW~Cnc and EX~Cnc.     
The \dss\ pulsations in these stars 
were discovered by \cite{gill91} and pulsation modelling 
was done in detail by \cite{gill92}. 
The current data set has a significantly higher signal-to-noise (S/N)
than the previous studies \citep{gill92, zhang05} and a superior spectral window.
We have for the first time unambiguously detected a very large number
of frequencies, comparable in number to the recent ambitious campaigns on
field \dss\ stars like FG~Virginis \citep{breger2005}.
We characterize EW~Cnc and EX~Cnc using 
a grid of seismic models, but unlike \cite{gill92} we 
also include the effects of rotation (in both equilibrium models and seismic
oscillations). 

Rotation will affect the internal structure of the stars, induce mixing
processes \citep{Zahn92, Heger00, reese06}, and cause significant 
asymmetric splitting of multiplets even in these
moderately rapid rotating stars \citep{Saio81, DG92, Soufi98, Sua06rotcel}.

\section{Blue straggler targets}

\begin{figure}
\includegraphics[width=8.8cm]{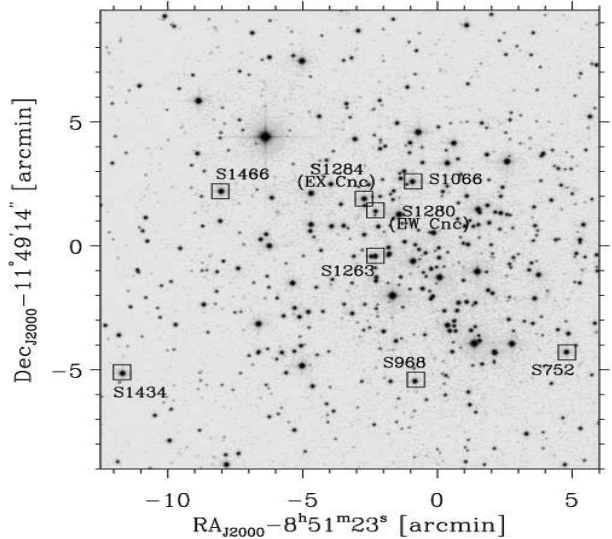}
\caption{Finding chart for the blue stragglers we have analysed with ID numbers from Sanders (1977).
The image is from the STScI Digitized Sky Survey.
\label{fig:find}}
\end{figure}

The locations of EW~Cnc and EX~Cnc in the colour-magnitude diagram
of \msyv\ are given in Fig.~\ref{fig:cmd}.
We also analysed six other BS stars inside the instability strip
and they are marked by filled circles. 
The $B-V$ colours and $V$ magnitudes are taken from \cite{mont93}. 
Only stars observed during the present campaign are plotted.
We have plotted an isochrone taken from the BaSTI database \citep{piet04}
for an age of 4.0\,Gyr with composition $Z=0.0198$, $Y=0.2734$,
adopting a distance modulus of $(m-M)_V=9.7$.
The instability strip is marked by dashed lines and is 
taken from \cite{breger01}. It has been transformed 
from the \str\ to the Johnson system using the calibration in \cite{cox00}. 
A finding chart showing the locations the BS targets
is given in Fig.~\ref{fig:find}.

\begin{figure*}
 \includegraphics[width=16cm]{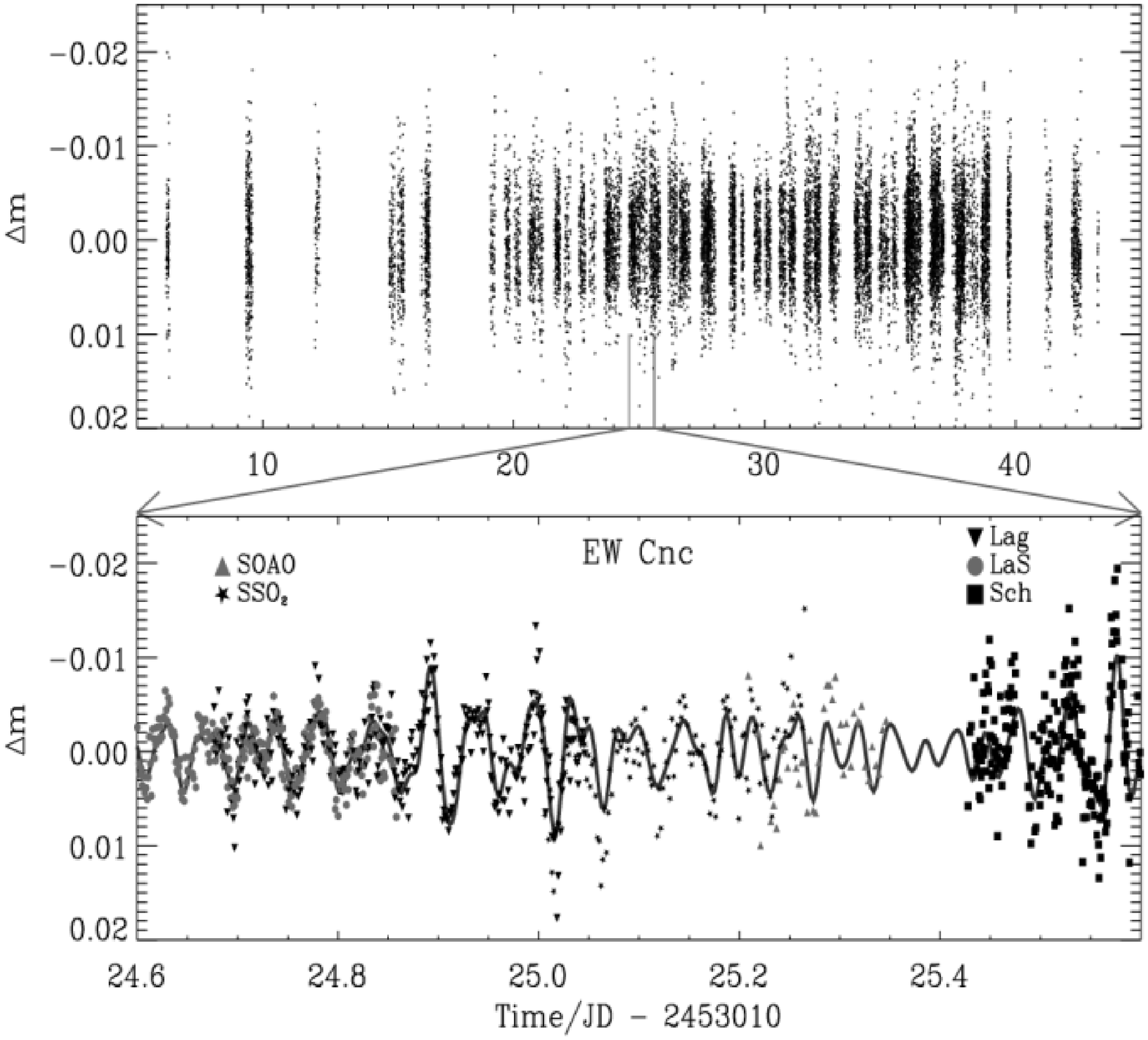}
\vskip -0.6cm
\hskip 0.25cm \includegraphics[width=16.0cm]{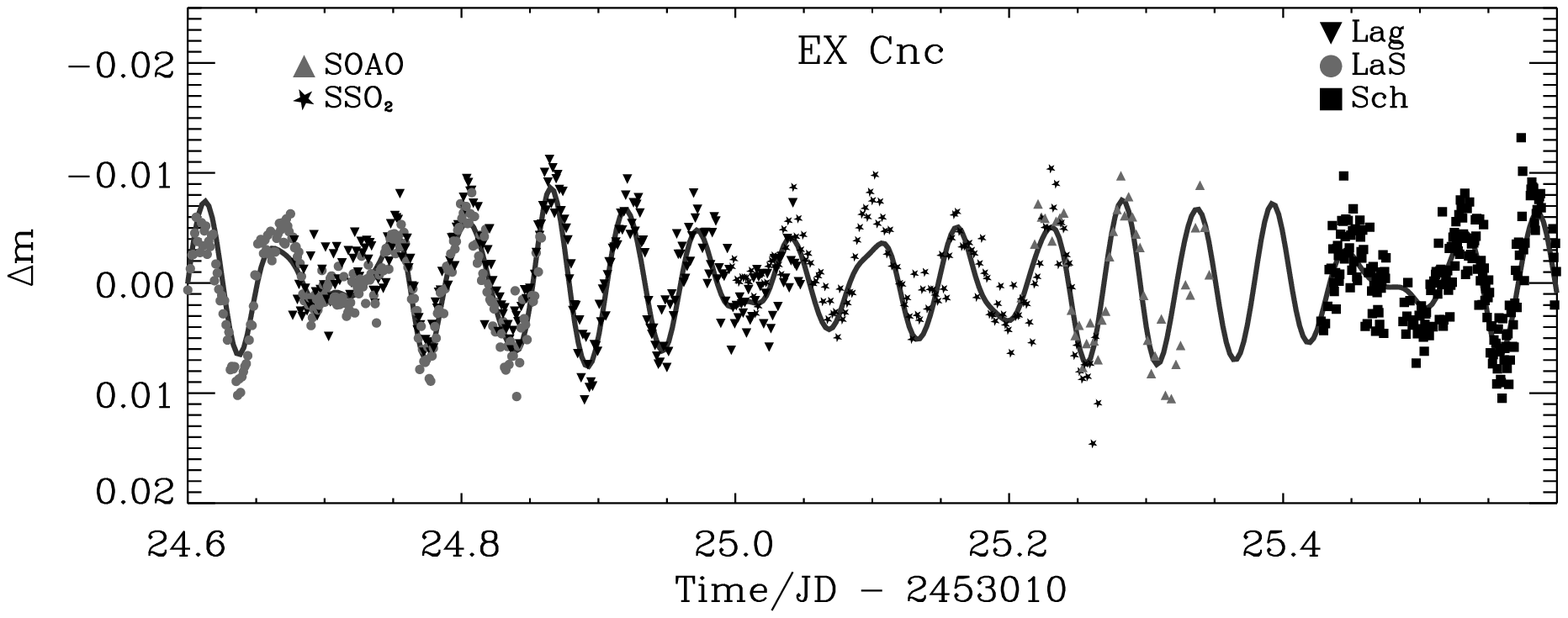}
 \caption{The complete light curve of EW~Cnc is shown in the \topp\ \panel.
The details of 24 hours with nearly full coverage 
are shown in the two \bott\ \panels\ for EW~Cnc and EX~Cnc where
different symbols are used for each observing site.
The continuous curves are the fits to the light curves.
\label{fig:lc}}
\end{figure*}

\section{Observations}

The data were collected in 2004 from January 6 to February 17 during
a multisite observing campaign with nine 0.6-m to 2.1-m class telescopes.
The photometric reduction was done using the \momf\ photometric package
\citep[Sects.~5--6]{kjeldsen92}.
The details of the observations and photometric data reduction 
were given in Paper~I.  

The choice of filters was made to optimize the signal-to-noise for the
observations of the giant stars in \msyv\ (see Paper~II). 
All sites used the
Johnson $V$ filters except Kitt Peak, where a Johnson $B$ filter was used. 
The bright BS star EX~Cnc was saturated in about half of the $B$ images.
We have a total of 450 hours of observation and about 15\,000 data points for each star.
The typical point-to-point scatter for EW~Cnc and EX~Cnc varies greatly from site to site,
and lies in the range 1--8 mmag. 
By comparison, \cite{gill92} used 37 hours of observation to obtain 2\,458 CCD frames
with 2 mmag point-to-point noise, which is similar to our best sites.

The complete light curve of EW~Cnc is shown in the \topp\ \panel\ in Fig.~\ref{fig:lc}. 
The \midd\ and \bott\ \panels\ show the details of the light curves 
of EW~Cnc and EX~Cnc during 24 hours. 
We have used different symbols for the five sites contributing 
to the light curve, as indicated in Fig.~\ref{fig:lc}.
The abbreviations for the five observatories are the same as in Paper~I. 
The solid curves are the final fits to the complete data set (see Sect.~\ref{sec:finfit}).

\section{Time-series analysis}

\subsection{Data point weights\label{sec:offsets}}

As can be seen qualitatively in Fig.~\ref{fig:lc} and as 
described in detail in Paper~I 
the quality of the photometry
varied a lot from site to site, in many cases from night to night,
and sometimes also during the night at a given site. 
To optimally extract the pulsation frequencies from 
the time series we have assigned weights to each data point. 

We calculated two kinds of weights, ``scatter weights'' ($W_{\rm scat}$) that are
based on the scatter in the time series and ``outlier weights'' ($W_{\rm out}$),
which suppress extreme data points.
The weights were calculated for each telescope and each star.
The scatter weights were determined from the average point-to-point 
scatter in a large ensemble of reference stars,
as described in detail in Appendix~\ref{sec:ptp}.
To calculate the outlier weights, we used the deviation
from zero of both the target star and the reference stars, 
as described in Appendix~\ref{sec:out}.

\begin{figure*}
\includegraphics[width=8.8cm]{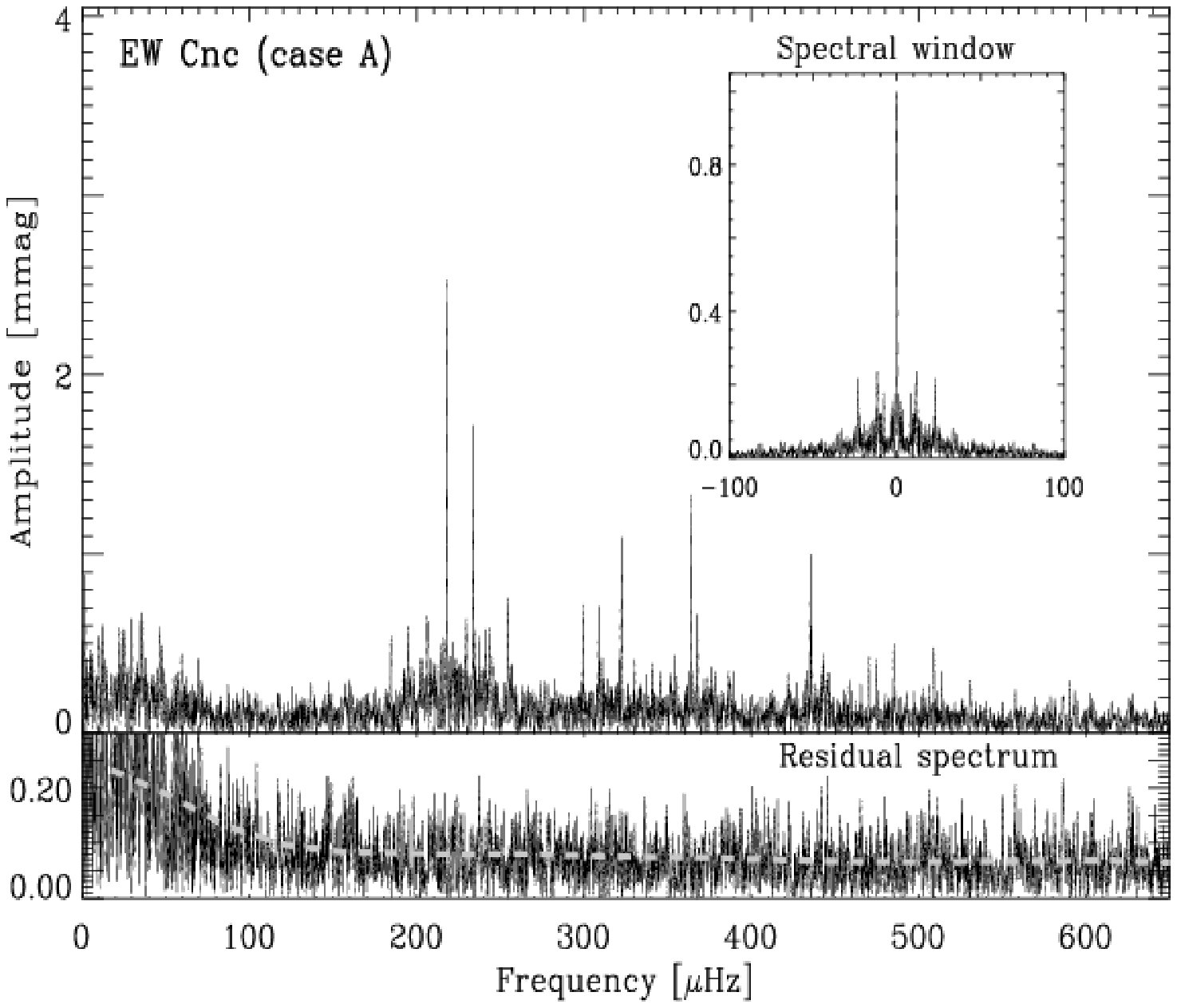}
\includegraphics[width=8.8cm]{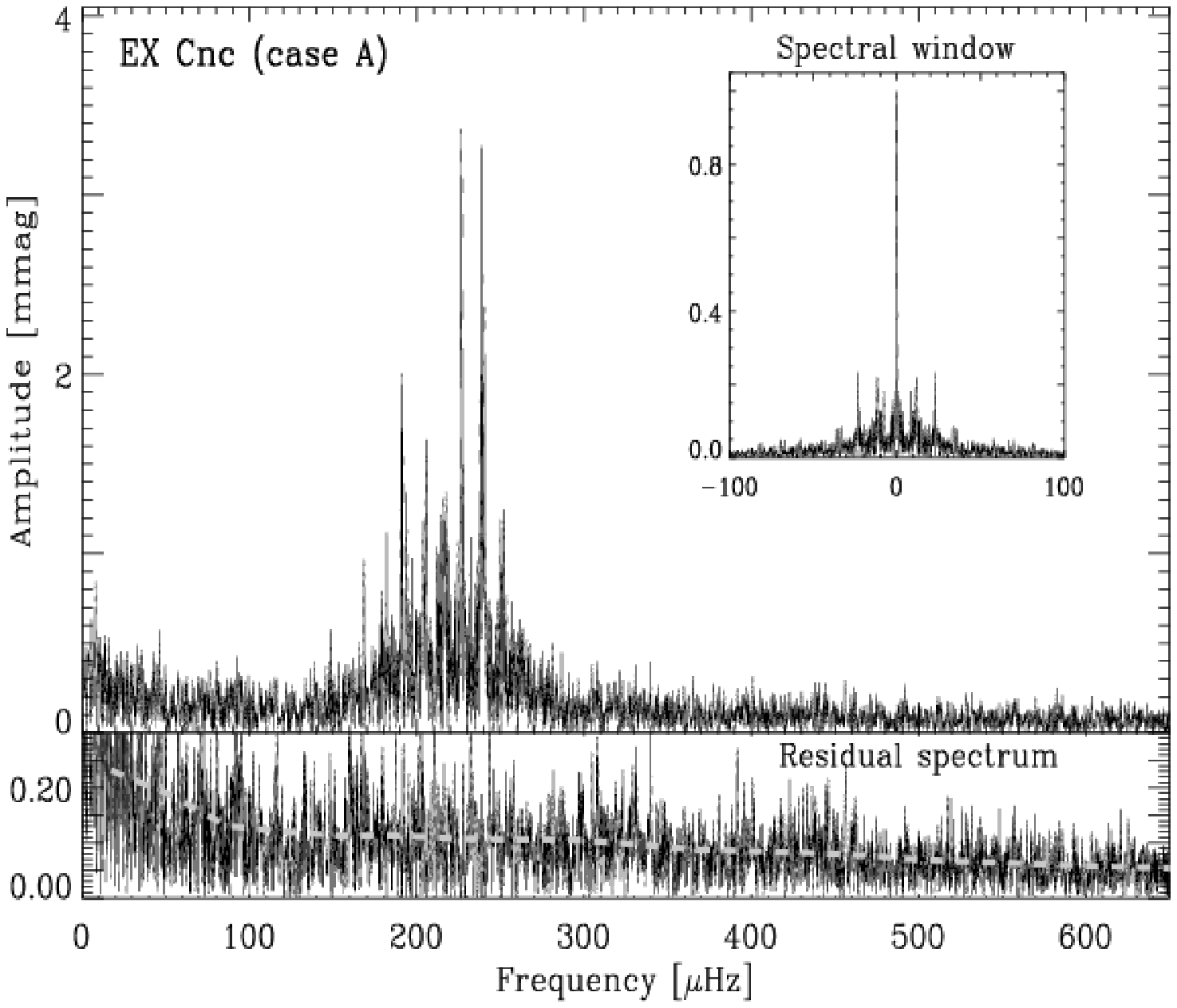}
\includegraphics[width=8.8cm]{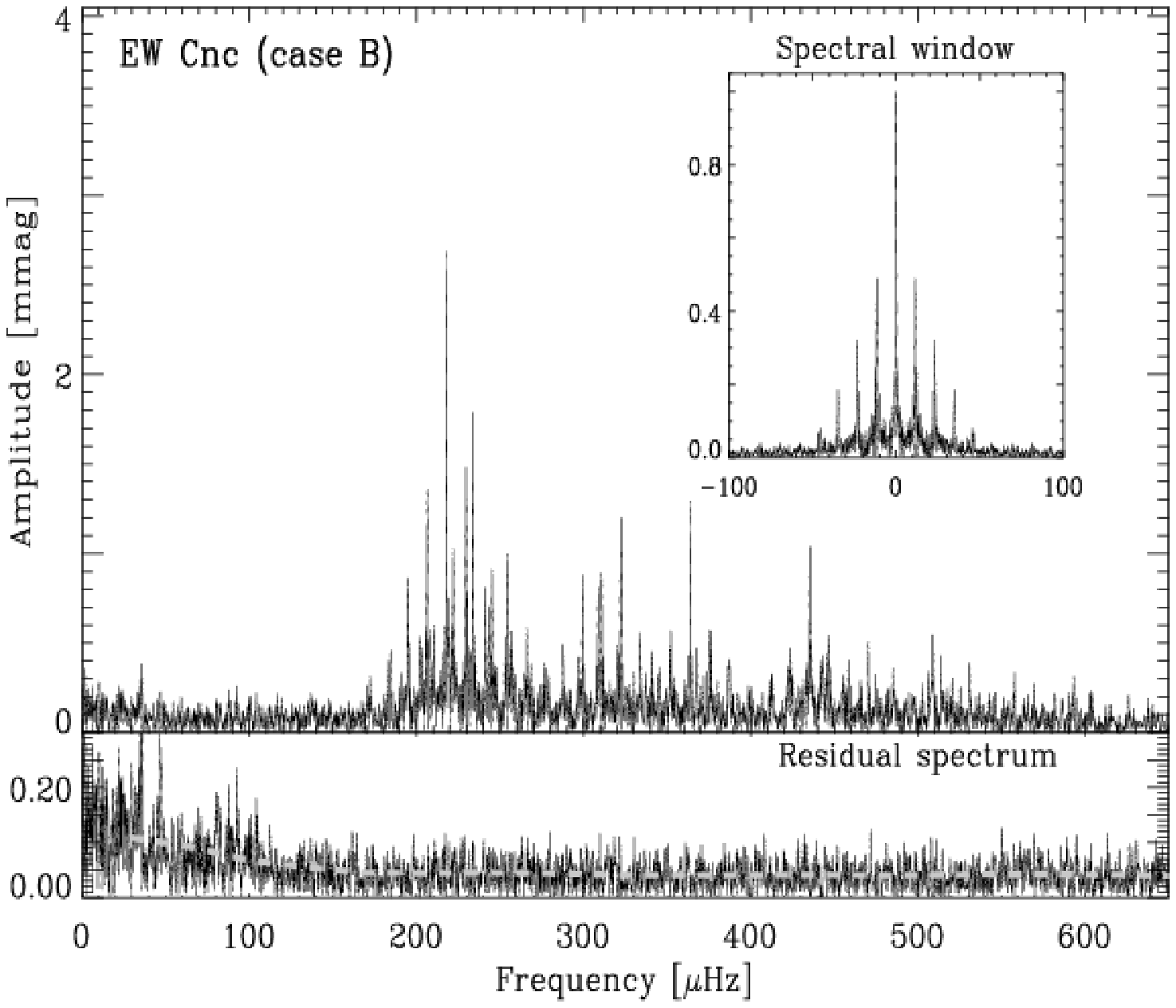}
\includegraphics[width=8.8cm]{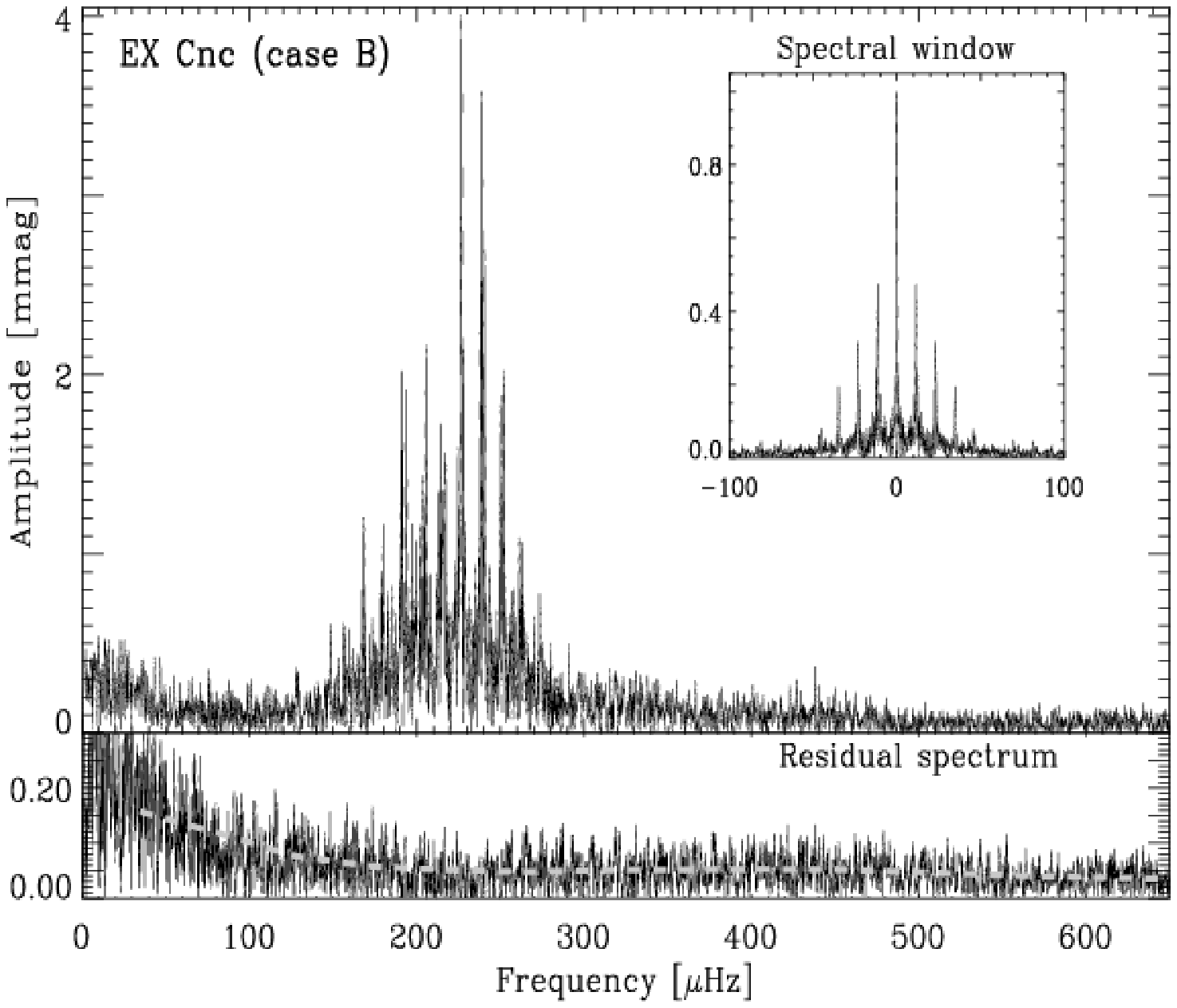}
\caption{Amplitude spectra of EW~Cnc (\lee\ \panels) and EX~Cnc (\rii\ \panels) for the
binned light curves (case~A, \topp\ \panels)
and the noise optimized light curves (case~B, \bott\ \panels).
The residual spectrum is shown below each plot and 
the insets show the spectral windows.
In the residual spectra the dashed grey lines mark the average noise level.
\label{fig:amp}}
\end{figure*}

%
\begin{table}
\caption{For case~A and B we give the noise level in the 
amplitude spectrum at high frequencies ($760\pm60$\,\mhz)
and the height of the first alias peak in the spectral window.
The results for our preferred cases are printed in italics.
\label{tab:noise}}      
\centering
\begin{tabular}{r|cccccc}
   \hline\hline  
     & \multicolumn{2}{c}{Case~A}                  & & & \multicolumn{2}{c|}{Case~B}           \\
     &       $A_{760}$            &     Alias      & & &  $A_{760}$               &    Alias   \\
Star &       [$\mu$mag]           &    [\%]        & & & [$\mu$mag]               &   [\%]     \\
   \hline  		      	     		       				    		  
EW~Cnc     &             55             &       23       & & & {\em   37   } & {\em   48 } \\ 
EX~Cnc &     {\em    48   }         & {\em  21  }    & & & {\rm   36   } & {\rm   47 }  \\ 
   \hline \hline                  
\end{tabular}
\end{table}
%

\subsection{Combining observations in $B$ and $V$}

At Kitt~Peak we used a Johnson $B$ filter while all
other sites used the Johnson $V$ filter.
The Kitt~Peak data were collected simultaneously with 
several other sites (see Fig.~2 in \cite{stello06}) but
since the precision is far superior we decided to include it.
The $B$ filter data comprise 1459 data points (9\%) for EW~Cnc
and 598 points (4\%) for EX~Cnc.

When combining the data we must correct for the dependence 
of the pulsation amplitude on the filter by scaling the $B$ filter data.
The scaling was found by measuring the amplitude ratios of the seven 
dominant frequencies in EW~Cnc when using the $B$ and $V$ filters alone.
The mean amplitude ratio was $0.74\pm0.02$\footnote{\cite{koen} observed 
two single-mode \dss\ stars (HD~199434 and HD~21190) in
the $B$ and $V$ filters and found amplitude ratios of $0.70\pm0.01$ and $0.70\pm0.03$, respectively.
We note that the empirical scaling is close to the ratio of the central wavelengths, 
\ie\ $\lambda_B/\lambda_V=434$\,nm$/\,545$\,nm\,$\simeq0.80$.}, which is the scaling we applied.
We also scaled the scatter weights found in Appendix~\ref{sec:ptp}
to ensure that the signal-to-noise is preserved.
The mean phase shift is $5.7\pm4.3^\circ$, 
which is not significantly different from zero.


\begin{table} 
\begin{center}
\caption{Frequency, amplitude, phase, and S/N of frequencies
detected in EW~Cnc for case~B. 
\label{tab:s1280}}
\begin{tabular}{r|lcl|r}
\hline\hline
ID        & \multicolumn{1}{c}{$f$ [$\mu$Hz]}  & $a$ [mmag] & \multicolumn{1}{c}{$\phi$} & S/N \\
\hline
$f_{1}$   &  217.993(4) &  2.66(2) &  0.05(1) &   48.0 \\
$f_{2}$   &  233.789(6) &  1.79(3) &  0.95(2) &   31.8 \\
$f_{3}$   &  363.811(4) &  1.29(3) &  0.00(1) &   26.9 \\
$f_{4}$   &  322.244(5) &  1.17(3) &  0.76(1) &   23.8 \\
$f_{5}$   &  435.39(1) &   0.94(4) &  0.19(3) &   20.8 \\
$f_{6}$   &  254.66(1) &   0.93(3) &  0.94(3) &   16.2 \\
$f_{7}$   &  308.81(1) &   0.75(2) &  0.66(3) &   14.7 \\
$f_{8}$   &  435.82(3) &   0.55(4) &  0.62(7) &   12.2 \\
$f_{9}$   &  367.83(1) &   0.58(3) &  0.93(3) &   12.2 \\
$f_{10}$  &  508.98(2) &   0.53(2) &  0.15(4) &   11.6 \\
$f_{11}$  &  443.07(3) &   0.46(3) &  0.78(7) &   10.1 \\
$f_{12}$  &  299.25(1) &   0.51(2) &  0.96(4) &    9.7 \\
$f_{13}$  &  470.16(2) &   0.41(3) &  0.93(5) &    9.1 \\
$f_{14}$  &  474.36(2) &   0.40(3) &  0.34(5) &    8.8 \\
$f_{15}$  &  513.75(2) &   0.40(2) &  0.50(5) &    8.6 \\
$f_{16}$  &  184.69(3) &   0.46(3) &  0.47(7) &    8.6 \\
$f_{17}$  &  485.12(2) &   0.38(3) &  0.76(6) &    8.5 \\
$f_{18}$  &  442.59(3) &   0.37(4) &  0.64(7) &    8.2 \\
$f_{19}$  &  530.87(3) &   0.36(2) &  0.62(7) &    7.5 \\
$f_{20}$  &  375.96(3) &   0.33(3) &  0.09(7) &    7.0 \\
$f_{21}$  &  354.22(3) &   0.33(3) &  0.92(9) &    6.9 \\
$f_{22}$  &  351.48(7) &   0.33(4) &  0.00(6) &    6.8 \\
$f_{23}$  &  202.14(3) &   0.37(3) &  0.06(7) &    6.7 \\
$f_{24}$  &  441.59(2) &   0.30(3) &  0.98(6) &    6.7 \\
$f_{25}$  &  557.87(3) &   0.32(3) &  0.15(8) &    6.7 \\
$f_{26}$  &  445.43(3) &   0.29(2) &  0.23(9) &    6.5 \\
$f_{27}$  &  593.32(3) &   0.30(2) &  0.82(7) &    6.2 \\
$f_{28}$  &  506.27(4) &   0.28(2) &  0.72(9) &    6.0 \\
$f_{29}$  &  311.87(3) &   0.29(2) &  0.13(8) &    5.7 \\
$f_{30}$  &  590.05(4) &   0.27(2) &  0.5(1)  &    5.7 \\
$f_{31}$  &  586.55(5) &   0.26(3) &  0.5(1)  &    5.5 \\
$f_{32}$  &  337.43(4) &   0.25(3) &  0.8(1)  &    5.2 \\
$f_{33}$  &  329.42(2) &   0.25(3) &  0.95(7) &    5.1 \\
$f_{34}$  &  586.15(5) &   0.24(3) &  0.1(1)  &    5.1 \\
$f_{35}$  &  400.09(5) &   0.24(3) &  0.4(1)  &    5.1 \\
$f_{36}$  &  235.16(4) &   0.28(3) &  0.3(1)  &    4.9 \\
$f_{37}$  &  475.67(5) &   0.20(2) &  0.5(1)  &    4.5 \\
$f_{38}$  &  413.53(6) &   0.20(3) &  0.7(1)  &    4.4 \\
$f_{39}$  &  625.85(4) &   0.21(2) &  0.8(1)  &    4.3 \\
$f_{40}$  &  315.65(4) &   0.20(2) &  0.3(1)  &    4.1 \\
$f_{41}$  &  421.70(4) &   0.19(2) &  0.8(1)  &    4.1 \\
\hline
\end{tabular}
\end{center}
\end{table}

To make sure that our results are not affected by the inclusion of the scaled $B$ data,
we repeated the analysis described below when using only the $V$ data.
For both stars we found exactly the same frequencies except for two frequencies in EW~Cnc
and three frequencies in EX~Cnc.
The most significant of these frequencies has S/N\,$=5.2$ while
the others have S/N below $4.4$.
The slight differences we find in the frequencies and amplitudes are
within the uncertainties found in Appendix~\ref{sec:vogtere}. 
We are therefore confident about 
combining the $V$ data with the scaled $B$ data.



\subsection{The optimal amplitude spectrum\label{sec:finalwei}}

We calculated the amplitude spectrum both
with the optimal spectral window and with the lowest noise level 
(see Kjeldsen et al.\ 2005). 
To do this we made the time-series analysis for two different approaches:

\begin{description}

\item[Case~A:] We computed the average light curve by 
binning all data collected within 8-minute intervals.
In the binning process we took into account the different 
quality of the data points by using weights $\wscat \cdot \wout$
for $n=2$ in Eq.~\ref{eq:ptpwei}. We did not use weights
when computing the amplitude spectrum a second time.

\item[Case~B:] No binning, but each data point
was given the weight $\wscat \cdot \wout$ 
when calculating the amplitude spectrum.

\end{description}

It was case~A that provided the optimal spectral window.
The amplitude spectra of EW~Cnc and EX~Cnc for case~A are shown in 
the \topp\ \panels\ in Fig.~\ref{fig:amp}. 
The residual spectrum after extracting the oscillation frequencies (see Sect.~\ref{sec:finfit})
is shown in each panel and the insets show 
the spectral windows, which have sidelobes at $\pm1$ \cday\ and $\pm2$ \cday\ of $\simeq23$\%.

The lowest noise level was found in case~B.
We tried both $n=1$ and $2$ in Eq.~\ref{eq:ptpwei} but
found that for $n=1$ the noise level was $5$--$10$\% lower, and
in addition the $\pm1$ \cday\ sidelobes were slightly lower. 
The fact that we found the lowest noise level for $n=1$ 
indicates that $1/(\wscat \cdot \wout)$ is the best
approximation to the true variance in the time series. 
We note that this was also concluded by \cite{handler03}.
The amplitude spectra for case~B are shown in the \bott\ \panels\ in Fig.~\ref{fig:amp}.  

While the noise level is significantly lower for case~B, 
the spectral window has much lower sidelobes in case~A.
In Table~\ref{tab:noise} we compare the noise levels and aliases. 
The noise was calculated
in a frequency band of $760\pm60$\,\mhz\ in the residual spectra.

In Appendix~\ref{sec:vogtere} we have made simulations of the time series to
check whether case~A or B gives the most robust results.
The conclusion is that case~A is preferred for EX~Cnc due 
to the high number of frequencies found in a relatively narrow interval, 
while we use case~B for EW~Cnc. 
The results for these cases are printed in italics in Table~\ref{tab:noise}.

\begin{table}
\begin{center}
\caption{Frequency, amplitude, phase, and S/N of frequencies
detected in EX~Cnc for case~A.
\label{tab:s1284}}
\begin{tabular}{r|lcl|r}
\hline\hline
ID        & \multicolumn{1}{c}{$f$ [$\mu$Hz]} & $a$ [mmag] & \multicolumn{1}{c}{$\phi$} & S/N \\
\hline
$f_{1}$   &  226.910(4)&   3.87(6) &  0.94(1) &   36.2 \\
$f_{2}$   &  238.883(5)&   3.74(6) &  0.60(1) &   35.0 \\
$f_{3}$   &  240.297(8)&   2.31(7) &  0.97(2) &   21.7 \\
$f_{4}$   &  191.464(9)&   2.08(7) &  0.45(2) &   18.3 \\
$f_{5}$   &  226.45(1) &   1.79(8) &  0.28(3) &   16.7 \\
$f_{6}$   &  205.49(1) &   1.68(8) &  0.14(3) &   15.2 \\
$f_{7}$   &  228.66(2) &   1.33(8) &  0.57(4) &   12.4 \\
$f_{8}$   &  215.33(1) &   1.34(8) &  0.41(4) &   12.3 \\
$f_{9}$   &  196.96(1) &   1.18(7) &  0.71(4) &   10.5 \\
$f_{10}$  &  190.92(1) &   1.16(8) &  0.46(3) &   10.2 \\
$f_{11}$  &  217.43(2) &   1.05(8) &  0.28(5) &    9.6 \\
$f_{12}$  &  211.82(2) &   1.04(7) &  0.77(4) &    9.5 \\
$f_{13}$  &  196.15(2) &   0.83(7) &  0.03(4) &    7.4 \\
$f_{14}$  &  193.93(2) &   0.79(7) &  0.11(6) &    7.0 \\
$f_{15}$  &  219.95(2) &   0.75(7) &  0.89(6) &    6.9 \\
$f_{16}$  &  236.34(2) &   0.60(5) &  0.04(7) &    5.6 \\
$f_{17}$  &  203.19(3) &   0.62(7) &  0.64(8) &    5.6 \\
$f_{18}$  &  223.01(4) &   0.59(7) &  0.19(9) &    5.5 \\
$f_{19}$  &  258.86(3) &   0.58(6) &  0.27(8) &    5.5 \\
$f_{20}$  &  199.45(3) &   0.59(7) &  0.76(7) &    5.3 \\
$f_{21}$  &  234.39(4) &   0.55(6) &  0.5(1) &    5.2 \\
$f_{22}$  &  229.14(5) &   0.55(5) &  0.5(1) &    5.2 \\
$f_{23}$  &  250.65(2) &   0.48(8) &  0.85(6) &    4.5 \\
$f_{24}$  &  256.86(3) &   0.47(6) &  0.55(9) &    4.4 \\
$f_{25}$  &  182.09(3) &   0.50(6) &  0.69(8) &    4.4 \\
$f_{26}$  &  148.54(3) &   0.49(6) &  0.20(7) &    4.3 \\
\hline
\end{tabular}
\end{center}
\end{table}

\subsection{Fourier analysis of EW~Cnc and EX~Cnc\label{sec:finfit}}

For each site the light curves were high-pass filtered to remove slow trends.
In this way we suppressed variations with frequencies below 80\,\mhz, corresponding
to periods longer than $\simeq3.5$ hours.
The pulsation frequencies are found mainly in the intervals from
200--600\,\mhz\ and 150--350\,\mhz\ for EW~Cnc and EX~Cnc, respectively.
We used the \period\ package by \cite{lenz05} to fit
the light curves in the frequency range 100--1000\,\mhz, as follows.
The amplitude spectrum was calculated using the light curve from 
case~B for EW~Cnc and case~A for EX~Cnc (see Sect.~\ref{sec:finalwei}).
The highest peak was then selected and the frequency, amplitude and phase were fitted.
This was done iteratively while in each step always improving the
frequencies, amplitudes and phases of previously extracted frequencies.

We extracted 41 and 26 frequencies with S/N\,$>4$ in the two stars,
and the parameters are given in Tables~\ref{tab:s1280} and \ref{tab:s1284}. 
The uncertainties are based on simulations, as described in Appendix~\ref{sec:vogtere}.
To determine the S/N of the extracted frequencies, we estimated the
noise level in the amplitude spectrum around each frequency. 
This was done after cleaning the amplitude spectra so only peaks 
with S/N\,$<3$ remained. 
The noise was calculated in bins with a width of 35\,\mhz\ and the noise estimate
is the average of the three nearest bins for each frequency.
The noise levels are marked by grey dashed lines 
in the residual spectra in Fig.~\ref{fig:amp}.

In comparison \cite{gill92} detected 10 frequencies in EW~Cnc and
6 frequencies in EX~Cnc.
We recovered these frequencies except one of their weaker modes at 
$423.2$\,\mhz\ in EW~Cnc and at $186.5$\,\mhz\ in EX~Cnc. 
\cite{zhang05} detected 4 frequencies in EW~Cnc and 
5 frequencies in EX~Cnc. They used data from a single 
observatory, and this explains why 
some of their frequencies are offset from ours by $\pm1$\,\cday.





\begin{figure*}
\hskip -0.3cm \includegraphics[width=18.4cm]{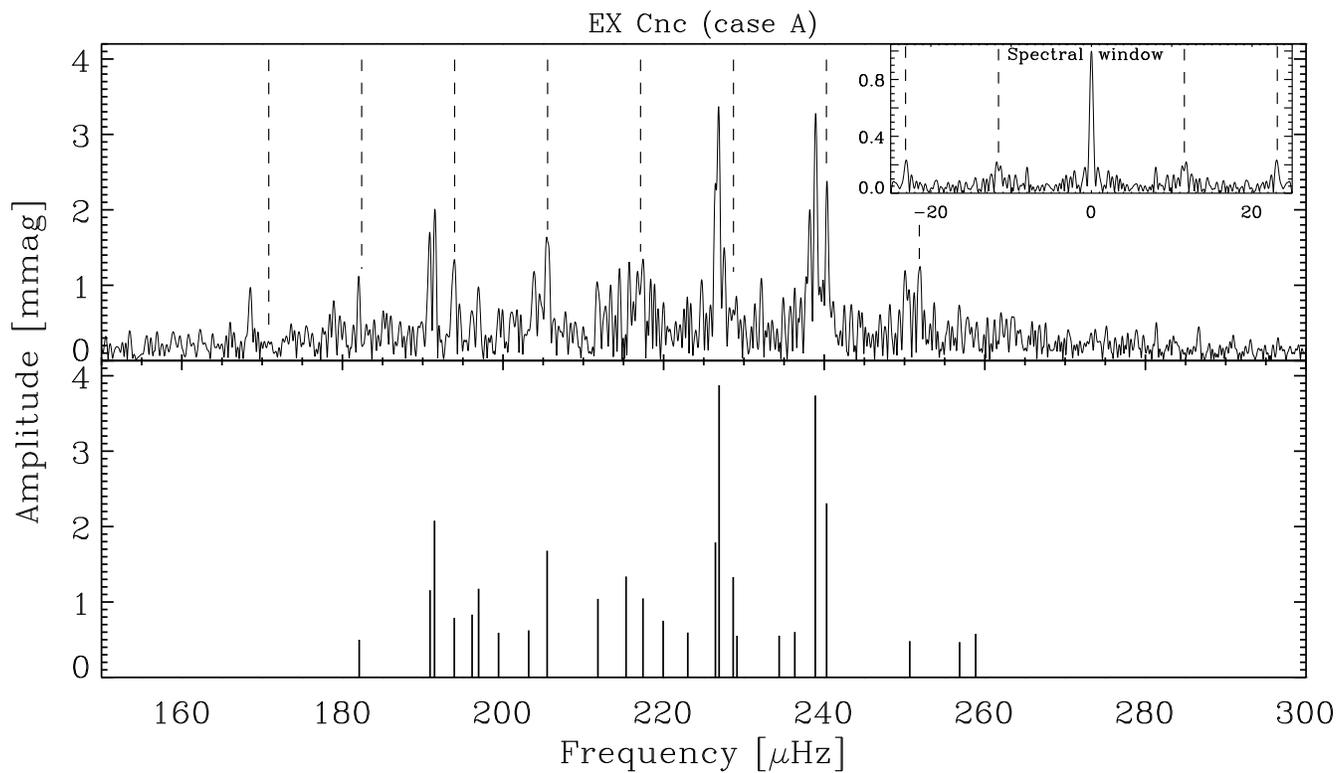}
\caption{The \topp\ \panel\ shows the amplitude spectrum of EX~Cnc for case~A.
The inset shows the spectral window on the same horizontal scale as the main Figure.
The vertical dashed lines are separated by 1\,\cday\ (11.574\,\mhz).
In the \bott\ \panel\ the detected frequencies are marked by vertical lines.
\label{fig:amp2}}
\end{figure*}

To check whether the frequencies we detected in the time-series
analysis were reliable, we made the time-series analysis 
for both cases A and B.
In both EW~Cnc and EX~Cnc we found that 
nearly all frequencies with S/N\,$>7$
were detected\footnote{The S/N was measured for case~B.} 
in both cases, while this was not true for the weaker frequencies.
Further, for the weaker frequencies that were detected in both
cases there was sometimes an offset of 1\,\cday.
In those cases, we chose the solution for case~A, 
which has the better spectral window.
As an additional check we used the 20 days in the time series 
with the best coverage and were able to recover all frequencies with S/N\,$>6$.
We also used only the data set from
La~Silla (14 consecutive nights of observation) but found
that for the weaker frequencies, the $\pm1$\,\cday\ alias was often
identified as the most significant frequency,
which is not surprising given the high sidelobes 
from the single-site time series.

Several of the 41 frequencies found in EW~Cnc are
found to be linear combinations;
examples are $f_{5} \simeq f_{13} \simeq 2 f_{1}$ and 
$f_{18} \simeq f_{5} - f_2 \simeq f_{12} - f_{28}$.
We searched for combinations within twice the 
resolution of the data set which is
$\Delta f$ = $2/T_{\rm obs}=0.63$\,\mhz. 
The condition is expressed as
$f_i \simeq n f_j + m f_k$ for any integer value $n,m \in [-2;2]$ and $i,j,k \in [1;41]$
and we found 13 combination frequencies.
We compared this with 1\,000 sets of 41 randomly distributed numbers
in the same interval as the observed frequencies, and we found
$18\pm6$ combinations. 
Hence our detection of 13 linear combinations in EW~Cnc is not surprising.
However, none of the frequencies found in EX~Cnc are
found to be linear combinations. This is also the case
for 1\,000 simulations of randomly distributed frequencies.
The reason for the different results for the two stars 
is that the frequencies in EX~Cnc are found 
in a much narrower frequency interval than for EW~Cnc.

\begin{table*}
\caption{Properties of the light curves and amplitude spectra of the observed BS stars in \msyv.
Star ID from Sanders (1977) and ID, $V$, and $B-V$ from Montgomery, Marschall, \& Janes (1993) are given.
$N$ is the number of data points.
The point-to-point scatter at the LaS and LOAO sites
and the noise level in the amplitude spectrum in the range $350\pm150$\,\mhz\ 
and $760\pm60$\,\mhz\ are listed. The last column contains membership probabilities from Sanders (1977).
\label{tab:bss}}      
\centering             
\begin{tabular}{llc|cc|rccrr|r}
   \hline\hline       
&  &     &       &        &        &  $\sigma_{\rm LaS}$ & $\sigma_{\rm LOAO}$ & \multicolumn{1}{c}{$A_{350}$}          & \multicolumn{1}{c}{$A_{760}$}  & \multicolumn{1}{r}{$P_{\rm Member}$}\\
Name & Sanders& MMJ  &   $V$ &  $B-V$ &  $N$   &  [mmag]             &  [mmag] & \multicolumn{1}{c}{[$\mu$mag]  }       & \multicolumn{1}{c}{[$\mu$mag]} & \multicolumn{1}{c}{[\%]}\\
\hline
               &S1466  & 6511 & 10.60 &  0.34  &  13316 &  1.7  &  3.3 &  66 &  41 &  0 \\     
               &S1434  & 6510 & 10.70 &  0.11  &   3312 &  $-$  &  3.3 & 372 & 160 & 91 \\     
EX~Cnc &S1284  & 6504 & 10.94 &  0.22  &  14660 &  1.9  &  2.9 & 240 &  65 & 95 \\     
               &S1066  & 6490 & 10.99 &  0.11  &  16954 &  1.7  &  2.6 &  46 &  35 & 90 \\     
               &S1263  & 6501 & 11.06 &  0.19  &  17408 &  4.6  &  2.7 &  89 &  69 & 89 \\     
               &S968   & 6479 & 11.28 &  0.13  &   9851 &  1.5  &  2.9 &  55 &  37 & 95 \\     
               &S752   & 6476 & 11.32 &  0.29  &   3316 &  $-$  &  2.8 & 194 & 120 & 95 \\     
 EW~Cnc    &S1280  & 5940 & 12.26 &  0.26  &  15524 &  1.9  &  3.3 & 188 &  46 & 93 \\     
   \hline \hline                  
\end{tabular}
\end{table*}

\subsection{Interpretation of the EX~Cnc spectrum\label{sec:interpret}}

In both cases A and B, we see a clear pattern of nearly equally spaced peaks
for EX~Cnc (Fig.~\ref{fig:amp}). 
In the \topp\ \panel\ in Fig.~\ref{fig:amp2} we show 
the amplitude spectrum of EX~Cnc (case~A) in more detail. 
To guide the eye we indicate a comb pattern with
a separation of $1\,$\cday\ (11.574$\,\mu$Hz) with dashed vertical lines. 
In the \bott\ \panel\ the frequencies detected 
in case~A (listed in Table~\ref{tab:s1284}) are marked by vertical lines.

We note that each dominant peak in the comb pattern consist of
several closely spaced peaks with remarkable similarity to what is seen for
stochastically excited and damped oscillations, also called solar-like
oscillations, which are driven by convection \citep{Anderson90}. 
We investigated whether the observed regular pattern and the closely spaced
lower amplitude peaks around each dominant peak could be explained by 
\begin{enumerate}
 \item Stochastically excited and damped oscillations of equally spaced modes.
 \item The spectral window (alias peaks).
 \item Closely spaced modes.
\end{enumerate}

(i) In $\delta\,$Scuti stars, the oscillations are caused by the
$\kappa$-mechanism (opacity driven) and are believed to be coherent, which
will produce a single isolated peak for each oscillation mode in the
amplitude spectrum. Although damped oscillations driven by convection might
be present in some $\delta\,$Scuti stars \citep{Samadi02}, the expected
frequencies and amplitudes are very different from what we see in
Fig.~\ref{fig:amp2}, which is consistent with opacity driven oscillations. 
The calculations by \citet{Samadi02} show frequencies in the range 400--1200$\,\mu$Hz and
amplitudes of roughly 100$\,$ppm for their selected models. 
These values are consistent with what we found specifically for this star 
using the parameters in Table~\ref{tab:fund}, a mass of 2\,$M_\odot$, 
and the usual scaling relations to predict the characteristics of
solar-like oscillations \citep{KjeldsenBedding95,Brown91}.
We find the central frequency (maximum amplitude)
to be $\nu_{\rmn{max}}\simeq 630\,\mu$Hz, amplitudes $\delta L/L\simeq 40\,$ppm, 
and mean frequency separation $\Delta\nu_{0}\simeq 40\,\mu$Hz). 
These properties of solar-like oscillations 
are clearly inconsistent with the observed amplitude spectrum.

(ii) Simulations showed that the regular pattern was unlikely to be
the result of only two or three modes. 
To reproduce the observed amplitude spectrum required at least 4--5 modes, 
equally spaced by roughly $1\,$\cday. The low alias sidelobes,
especially in case~A, imply that the pattern is intrinsic to the star. 
Many of the lower amplitude peaks around each dominant peak
seem to be separated by exactly 1 or $2\,$\cday\ from a neighbouring
peak. However, the strength of these lower amplitude peaks are in many
cases higher than expected from the strength of the sidelobes in the
spectral window.  

(iii) Closely spaced peaks are not uncommon in $\delta\,$Scuti stars
\citep{BregerBischof02}. 
\cite{bregerpamyat2006} noted that 18 pairs ($\Delta f<1$\,\mhz) 
exist in the \dss\ star FG~Virginis \citep{breger2005}. 
They found evidence for the frequency doublets being real.
As seen in Fig.~\ref{fig:amp2}, we also detect several close pairs.

In summary, damped oscillations driven by convection seems to be a very
unlikely cause for the characteristic pattern we see in Fig.~\ref{fig:amp2}. 
We believe the pattern is most likely due to a nearly regular series of at
least five modes (excited by the $\kappa$ mechanism) which, due to a
general spacing very close to $1\,$\cday, shows a large number of small peaks
around each  dominant peak caused by the spectral window. 
This is further enhanced by the presence of lower amplitude modes 
close to the dominant modes (see \bott\ \panel\ in Fig.~\ref{fig:amp2}).
Our interpretation is supported by \cite{gill92} who found
frequency separations of $\sim5$\,\mhz\ and $\sim10$\,\mhz\ in this star,
and suggested that the latter might be due to rotational splitting.

\subsection{A search for pulsations in six other BS stars\label{sec:bss}}

In addition to EW~Cnc and EX~Cnc, 
we observed six other BS stars
in the instability strip as shown in Fig.~\ref{fig:cmd}.
To search for pulsations in the stars we calculated
the amplitude spectra in the range 0--1000\,\mhz, 
using weights as for case~B (Sect.~\ref{sec:finalwei}).
In Table~\ref{tab:bss} we summarize the properties
of the BS stars observed during the campaign. 
For completeness, EW~Cnc (S1280) and EX~Cnc (S1284) are included in the Table.
We give the ID numbers from \cite{sanders77} and \cite{mont93}, and the
$V$ magnitude, and $B-V$ colour from the latter. 
The total number of data points $N$ and \isd\ (Eq.~\ref{eq:isd})
for the La~Silla and LOAO (Mt.\ Lemmon Optical Astronomy Observatory) 
sites indicate the quality of the data.
We note that La~Silla had very good weather conditions 
during the campaign, while LOAO had worse conditions (for more details see Paper~I). 
The mean level in the amplitude spectrum in the
range where oscillations are expected is denoted $A_{350}$
($200$--$500$\,\mhz), 
while that at slightly higher frequencies $A_{760}$ ($700$--$820$\,\mhz)
is a measure of the white noise.
In the last column, the membership probability from \cite{sanders77} is given.
We note that more recent radial velocity studies 
by \cite{girard89} and \cite{zhao93} are in general agreement with this. 


Two of the BS stars (S1434 and S752) 
were only observed by a few sites. Hence, for these stars the
noise level in the amplitude spectrum is about a factor 4--6 higher than
for the other stars.
We find no significant peaks in the range 0--1000\,\mhz\ 
above S/N of 4\footnote{Defined as $4 \cdot A_{350}$ where $A_{350}$ is from Table~\ref{tab:bss}.}
in any of the BS stars except EW~Cnc and EX~Cnc.

Our results are in agreement with \citet{sandquist03a} who looked
for photometric variability in several BS stars in \msyv. Except for 
EW~Cnc and EX~Cnc they found no \dss\ pulsations in their sample, 
which included S752, S968, S1066, and S1263. The latter was also observed by
\citet{gill92} who placed an upper limit of variation at 0.2 mmag 
(corresponding to $\sim5\,$mmag scatter in the time series) for
frequencies above 130\,\mhz, in agreement with this study and 
\citet{sandquist03a}.
We note that \citet{stassun02} reported unusually high scatter 
of up to 30 mmag in this star from a photometric $BVI$ time-series study. 
The relatively high noise level for S1263 compared to the other BSs in
our data (see $\sigma_{\mathrm{LaS}}$ in Table~\ref{tab:bss}) 
is due to the star being blended (see Fig.~\ref{fig:find}).

\begin{table}
\begin{center}
\caption{Fundamental parameters of EW~Cnc and EX~Cnc.
 \label{tab:fund}}
\begin{tabular}{l|cc|cc}
\hline\hline
   & \multicolumn{2}{c|}{EW~Cnc}  & \multicolumn{2}{c}{EX~Cnc} \\
   & \teff\ [K] & \logg & \teff\ [K] & \logg \\
\hline
Mathys et al.\ 1991      & 8090 & 4.33 &  7750 & 3.79 \\
Gilliland \& Brown 1992  & 7960 & 4.20 &  7900 & 3.88 \\
Landsman et al.\ 1998    & 8090 & 4.12 &  7610 & 3.78 \\
$uvby$ ({\em this study})& 7800 &      &  7840 &      \\
Geneva ({\em this study})& 8080 & 4.59 &       &      \\

\hline
\end{tabular}
\end{center}
\end{table}

\section{Theoretical models of EW \& EX~Cnc}\label{sec:models}

To model EW~Cnc and EX~Cnc we used 
the stellar evolution code {\sc cesam} \citep{morel97}.
The effects of rotation are taken into account in the equilibrium
equations as a first-order perturbation to the local gravity, 
while assuming rotation as a rigid body.  
The {\sc ceff} equation of state was used \citep{cd92} 
and {\sc opal} opacities were adopted \citep{iglesias96}.
A more detailed description of the input physics 
for the {\sc cesam} code was given by \cite{casas06}. 

We used the oscillation code {\sc filou} \citep{filou,SuaThesis} 
to calculate the oscillation frequencies from the evolution models. 
We assume rigid-body rotation and 
the effects up to second order in the 
centrifugal and Coriolis forces were included. 
We note that \cite{reese06} found that this perturbation 
approach may be invalid even at moderate rotational velocities.
Effects of near degeneracy are expected to be significant (Soufi et al.\ 1998)
as was described in detail by \cite{Sua06rotcel}.


%
\begin{table}
\caption{Parameters of theoretical pulsation models for EW~Cnc and EX~Cnc, which 
are inside the photometric error boxes shown in Fig.~\ref{fig:models}.
For each mass the range of effective temperatures is given. 
For each star we computed models 
with two different rotational velocities given in the last column.
\label{tab:models}}
\centering
\begin{tabular}{c|cccc}
   \hline\hline       
Star &   $M/M_\odot$ & \teff~[K] & $v_{\rm rot}$ [\kms] \\
   \hline

EW~Cnc & $1.7$ & $7930$--$8110$ & $90, 150$ \\
       & $1.8$ & $7930$--$8210$ &    $-$ \\
       & $1.9$ & $7930$--$8210$ &    $-$ \\
       & $2.0$ & $7940$--$8200$ &    $-$ \\
			          
   \hline		          
			          
EX~Cnc & $1.8$ & $7450$--$7620$ & $65, 120$ \\
       & $1.9$ & $7450$--$7740$ &    $-$ \\
       & $2.0$ & $7480$--$7740$ &    $-$ \\
       & $2.1$ & $7470$--$7710$ &    $-$ \\
       & $2.2$ & $7480$--$7730$ &    $-$ \\
       & $2.3$ & $7630$--$7740$ &    $-$ \\

   \hline \hline                  
\end{tabular}
\end{table}
%

\subsection{Fundamental atmospheric parameters\label{sec:fund}}  

The fundamental atmospheric parameters of EW~Cnc and EX~Cnc 
were estimated by \cite{mathys91} and \cite{gill92} based on 
\str\ photometry by \cite{nissen87} but applying different calibrations \citep{moon85,philip79}.
We used the more recent calibration by \cite{napi93}, 
while assuming the mean cluster reddening 
from Nissen et al.\ (1987). 
For EW~Cnc we used Geneva photometry from
\cite{geneva} and applied the calibration by \cite{kunzli97}.
\cite{landsman98} constrained \teff\ of both stars based on UV measurements.
We summarize the results in Table~\ref{tab:fund}. 
Typical uncertainties on \teff\ and \logg\ are 150~K and 0.2 dex 
and there is good agreement between the different calibrations.

An important limitation of our modelling is the uncertainty on
the composition of the stars. The composition will
depend on the formation scenario of these BS stars, 
which we certainly cannot begin to constrain with the current data set.
\cite{gill92} discussed different formation scenarios for
the same stars from either a direct collision or gradual coalescence.
They made the important point that it
would be difficult to detect the effects of a $5$\% increase of 
helium, \eg\ as a result of mixing two evolved main sequence stars.
For simplicity we assume that the metallicity of EW~Cnc and EX~Cnc is 
the same as for the cluster. \cite{mont93} found \feh\, $=-0.05$ but 
they did not estimate the uncertainty; a conservative estimate is $\pm0.10$ dex.


\begin{figure}
\includegraphics[width=8.8cm]{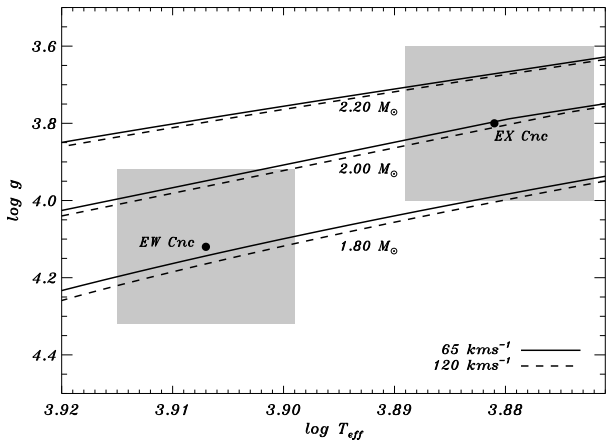}
\caption{The $\log g-$\teff\ diagram showing selected evolution tracks from our model grid.
The $1-\sigma$ photometric error boxes of EW~Cnc and EX~Cnc are indicated as grey squares.
\label{fig:models}}
\end{figure}

\begin{figure*}
\includegraphics[width=8.8cm]{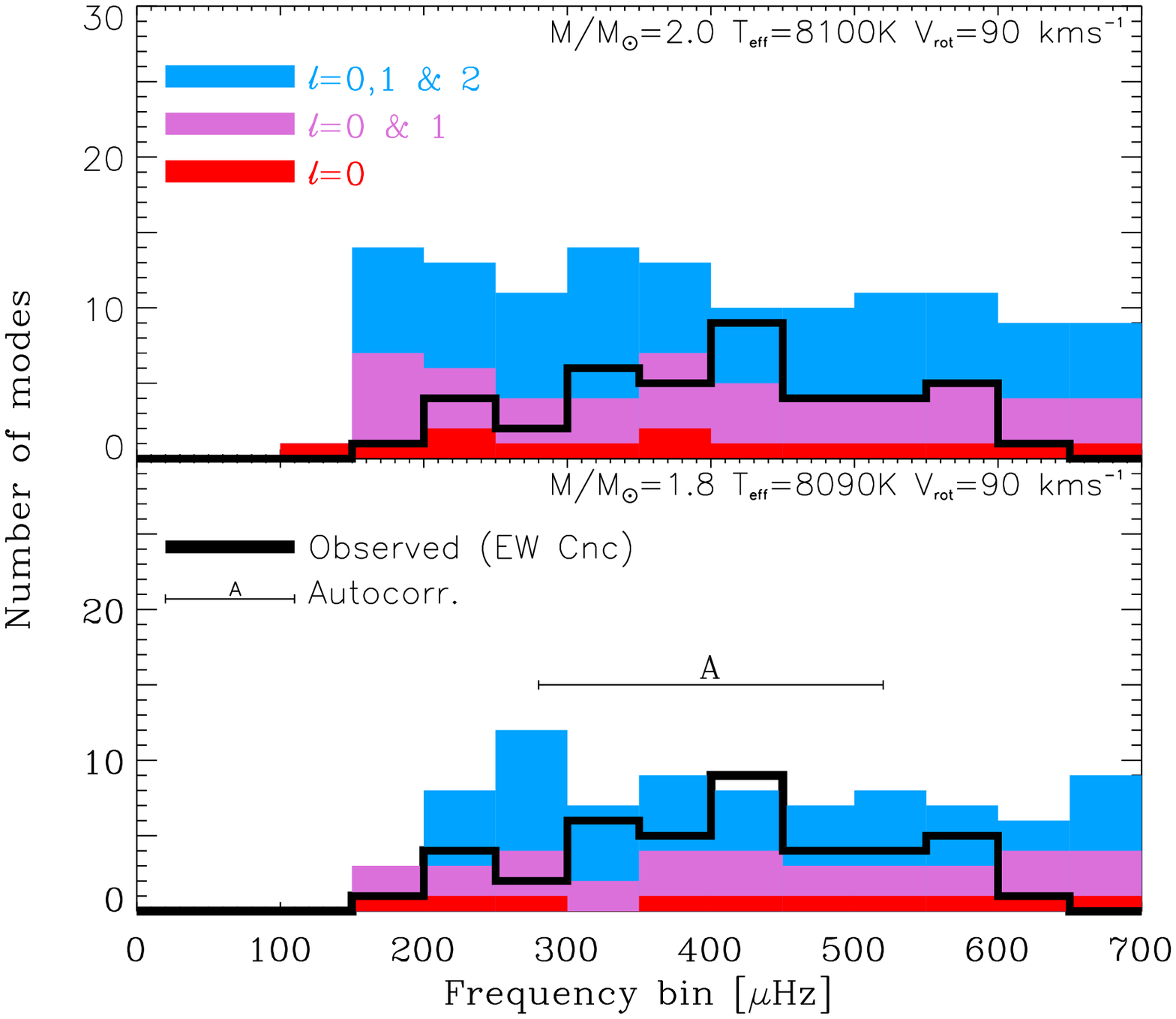}
\includegraphics[width=8.8cm]{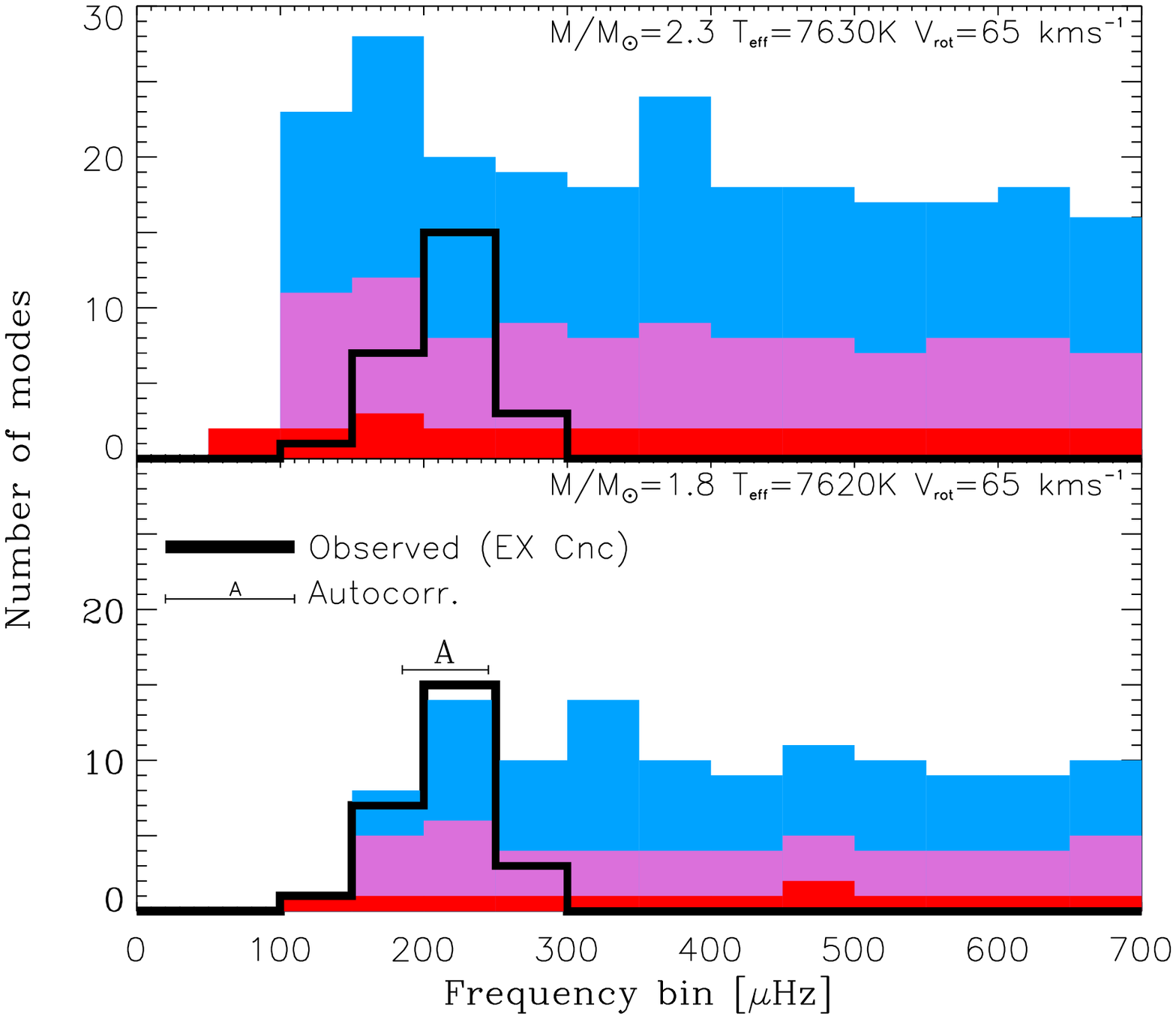}
\caption{Histograms showing the number of frequencies 
in EW~Cnc (\lee\ \panel) and EX~Cnc (\rii \panel) in frequency bins of 50\,\mhz. 
The filled histograms are for pulsation models and the colour describes how high angular degree is included.
The thick black lines are for the observations.
The \topp\ and \bott\ \panels\ are for pulsation models with high and low mass with the adopted \teff.
The models in the \bott\ \panels\ both have mass $1.8\,M_\odot$, 
slightly different $v_{\rm rot}$, and quite different \teff.
The horizontal bar marked ``A'' marks the region used for the autocorrelation in Sect.~\ref{sec:corr}.  
\label{fig:nmode}}
\end{figure*}

\subsection{The model grid\label{sec:grid}}        

We have calculated a grid of stellar models consistent
with the fundamental parameters of EW~Cnc and EX~Cnc.
A few evolution tracks are shown in Fig.~\ref{fig:models} in the $\log g$ \vs\ \teff-diagram, where
the solid and dashed lines correspond to rotational velocities of 65 and 120\,\kms.
The photometric error boxes are shown as grey squares.
The center of each box is the \teff\ and \logg\ from \cite{landsman98} 
and the half-widths are equal to the estimated uncertainties of 150\,K and 0.2 dex. 
We have computed pulsation frequencies for the models inside the photometric error boxes.
The mass ranges are $1.7$--$2.0$\,\msun\ for EW~Cnc and $1.8$--$2.3$\,\msun\ for EX~Cnc in steps of $0.1$\,\msun.
Further details of the grid of pulsation models are given in Table~\ref{tab:models}.

Analyses of stellar spectra \citep{peterson84, manteiga89, pritchet91} 
reveal moderate projected rotational velocities for EW~Cnc and EX~Cnc,
and we have used average values of \vsini\,$=90$ and $65$\,\kms, respectively.
However, the inclination angle is unknown, and therefore we 
calculated models and corresponding oscillation frequencies 
for the minimum and a moderate value of $v_{\rm rot}$ for each star.
The values in the last column in Table~\ref{tab:models}  
correspond to inclination angles of $i=90^\circ$ and $45^\circ$.


\subsection{Comparison of observed and theoretical frequencies\label{sec:corr}}

In Fig.~\ref{fig:nmode} we show the number of modes as a function of frequency, in bins of 50\,\mhz.
The black histograms are for the observed frequencies for EW~Cnc (\lee\ panels) and EX~Cnc (\rii\ panels), 
which are listed in Tables~\ref{tab:s1280} and \ref{tab:s1284}.
The solid histograms are for frequencies from pulsation models 
and the colours indicate how high angular degree, $l$, we have included\footnote{In the integrated 
light from the stars we expect to observe only low degree modes. 
Modes with $l\geq 3$ will have very low observed amplitudes 
due to geometrical cancellation effects when averaging over the stellar disc. 
We note that for each mode we include all $2l+1$ possible values of $m$.}.
The \topp\ and \bott\ \panels\ are for models with high and low mass.
The low-mass models both have a mass $1.8\,M_\odot$, while
the high-mass models correspond to the evolution tracks 
that graze the photometric error boxes,
\ie\ $2.0$ and $2.3\,M_\odot$ for EW~Cnc and EX~Cnc, respectively.
We used the \teff\ adopted for the centres of the 
photometric error boxes in Fig.~\ref{fig:models}. 

By inspecting the histograms for the models in Fig.~\ref{fig:nmode} 
one can see that the number modes increase 
when increasing the mass or decreasing \teff.       
Histograms for models with different rotational velocities are nearly 
identical since the frequency shifts are relatively small ($\simeq$\,few \mhz) 
for these moderate velocities.
The histograms of all pulsation models within the photometric error boxes 
are in agreement with the observations, but {\em only} when including modes with degrees $l=0,1,$ and $2$. 
However, if the masses are lower than $1.7$ and $1.8$\,$M_\odot$ for 
EW~Cnc and EX~Cnc, respectively, it will require lower temperatures, 
which are outside the photometric error boxes.
In other words, if a more accurate 
\teff\ becomes available we can put a lower limit on the mass.

In EW~Cnc the detected frequencies are evenly distributed over a wide interval,
while in EX~Cnc they are concentrated in a narrow band.
The horizontal bars labeled ``A'' in Fig.~\ref{fig:nmode} indicate where most observed frequencies are found.
In these frequency intervals we estimate that we have 
detected $40$--$60$\% of the modes in EW~Cnc and $70$--$100$\% in EX~Cnc. 
These results rely on the assumptions that the considered ranges in mass and \teff\ are realistic and
that only $l\leq2$ modes have detectable amplitudes.


\begin{figure*} 
\includegraphics[width=8cm]{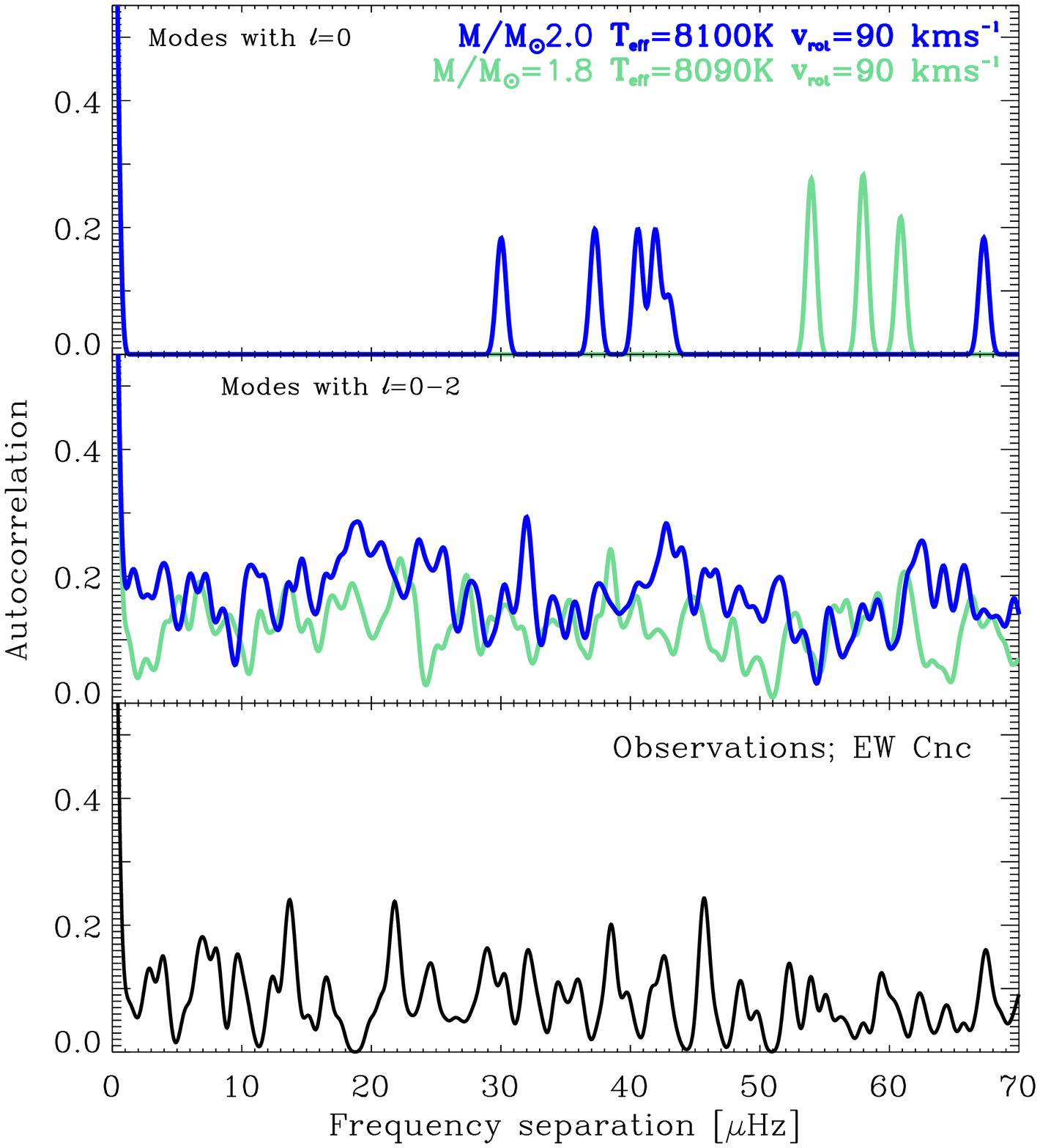}
\includegraphics[width=8cm]{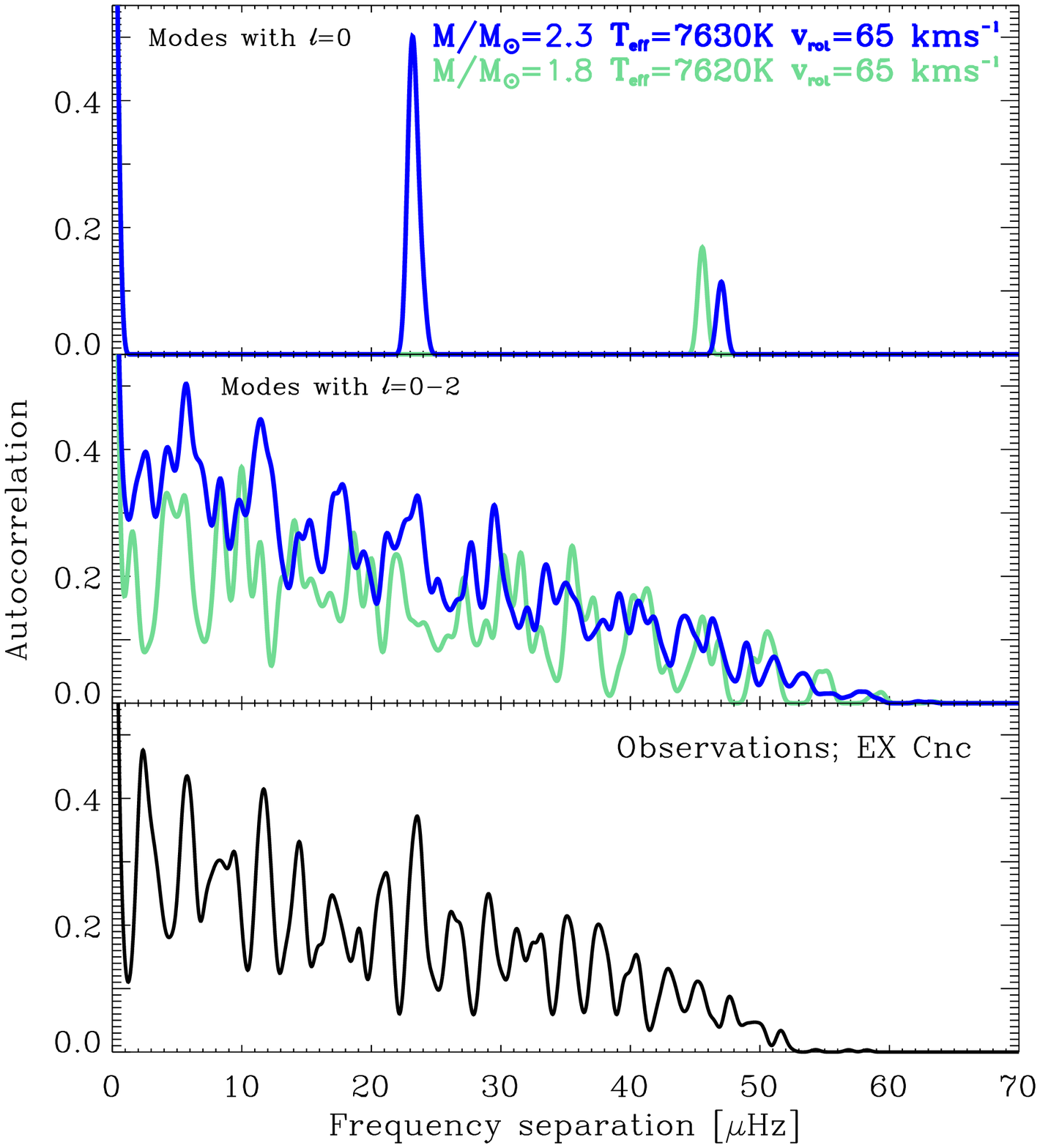}
\caption{Comparison of the autocorrelations of frequencies from pulsations
models and the observed frequencies in EW~Cnc (\lee\ \panels) and EX~Cnc (\rii \panels).
In the \topp\ \panels\ we included only radial modes while in
the \midd\ \panel\ we included modes with $l=0,1$ and $2$. 
We show results for the same models as in Fig.~\ref{fig:nmode}.
The \bott\ \panels\ are the autocorrelations of the observed frequencies.
\label{fig:autoexcnc}}
\end{figure*}

To further investigate the distribution of peaks and search for recurring patterns
we computed autocorrelations for the frequencies from the observations and the pulsation models.
This method can in principle be used to extract
important quantities for asteroseismic modelling 
like the large and small separation and rotational splitting.
This would enable us to constrain the mass and temperature of the stars.
Since we have no information about the relative amplitudes
of the theoretical frequencies, 
each frequency is represented by a Gaussian of unit height and
with a FWHM equal to twice the formal frequency resolution of the data set.
To be able to compare the autocorrelations of the observed and theoretical frequencies, 
we follow the same procedure for the observations,
\ie\ only using the frequencies and not the observed amplitudes.
We use the frequencies from the regions where 
the majority of the frequencies are observed. 
These regions are marked by horizontal bars labeled ``A'' in Fig.~\ref{fig:nmode}.
We tried different widths of these
bands, but the conclusion summarized here is the same.

In Fig.~\ref{fig:autoexcnc} we show autocorrelations for theoretical and observed
frequencies for EW~Cnc (\lee\ \panels) EX~Cnc (\rii \panels). 
The models have the same masses and temperatures as shown in the histograms in Fig.~\ref{fig:nmode}.
In the \topp\ panel only radial modes are included. 
The results for EX~Cnc shows the
large separation at $\simeq23$\,\mhz\ for the model with high mass, while 
for the less massive model there is a low peak at twice the large separation 
(see \topp\ \rii\ \panel\ in Fig.~\ref{fig:autoexcnc}). 
In the \midd\ \panel\ we included all model frequencies with $l=0,1$, and $2$; 
the higher number of modes makes the interpretation of the 
autocorrelation much more difficult. 
In the \bott\ \panels\ we show the autocorrelations of the observed frequencies.
There is an intriguing similarity between the autocorrelation
of the observed frequencies in EX~Cnc and the model with high mass 
(the \midd\ and \bott\ \rii\ \panels\ in Fig.~\ref{fig:autoexcnc}).
However, we are extremely cautious not to over-interpret 
the only marginally significant autocorrelation peaks at
2, 6, 12, and 24\,\mhz.

Although there appears to be qualitative agreement 
between the observations and the theoretical models, 
we cannot at this stage use the method to determine which combination of 
mass, evolutionary stage, or rotational velocity best
describes the observations. 
To make progress and make a detailed comparison of the observed 
frequencies with theoretical models would
require that we have identified at least some of the dominant modes.
This can be achieved from the study of line-profile variations
or from comparison of phase shifts of the frequencies in 
different photometric filters \citep{dasz03}.
We cannot apply the latter method since 
most of our observations were done in the $V$ band with only a 
few nights of data in $B$.

\section{Conclusion\label{sec:discuss}}


We have analysed a new photometric data set of eight BS stars in \msyv.
Only in the two known variable stars EW~Cnc and EX~Cnc do we find \dss\ pulsations. 
We detect 41 frequencies in EW~Cnc and 26 frequencies in EX~Cnc.
Compared to the previous multisite campaign by \cite{gill92} we have 
detected four times as many frequencies in these two stars, and 
the number of frequencies is comparable to the best-studied field \dss\ stars.

The frequencies with the highest amplitudes in EX~Cnc are separated by 
roughly 1\,\cday\ (see Fig.~\ref{fig:amp2}) and 
our extensive multisite data are therefore essential to study the pulsations in this star.
We have made simulations of the light curves which
show that the frequencies are recovered in all cases (see Appendix~\ref{sec:vogtere}).
We repeated the Fourier analyses on subsets of the data to confirm that 
the same frequencies were extracted.
Thus, we claim that the observed
pattern cannot be due to the spectral window, but represents 
a real property of the distribution of frequencies in EX~Cnc.
Additional support for this claim is the fact 
that a similar pattern is not seen in EW~Cnc.

We computed a grid of theoretical pulsation models 
that take the effects of rotation into account,
which were not included in earlier models \citep{gill92}.
We find general agreement between the observed and computed 
frequencies when comparing their distribution in histograms. 
We can claim to have detected non-radial modes and
in the range where most of the frequencies are observed,
and at least $40$\% and $70$\% of all low-degree ($l\leq2$) 
modes are excited in EW~Cnc and EX~Cnc, respectively.
This is perhaps the most important result of the current study, since
in the past decade of observational work on \dss\ stars 
only a tiny fraction of the frequencies expected
from theoretical models were in fact detected.
A reasonable explanation is that high S/N and 
good frequency resolution is needed to be able 
detect the weakest modes as was suggested by \cite{breger2005} and \cite{bruntt07}.

The autocorrelation of the observed frequencies 
does not show any significant peaks in any of the stars.
However, autocorrelations of the frequencies from theoretical pulsation 
models show that if indeed modes with degree $l=0,1$, and $2$ are observed, 
we cannot expect to unambiguously detect the large separation from such an analysis. 
This result was obtained while assuming that all amplitudes 
are equal, since no amplitude information is available for the theoretical models.

To improve on the interpretation of the rich 
frequency spectra of EW~Cnc and EX~Cnc, 
and other well-observed \dss\ stars in general, we see two possibilities:
\begin{itemize}

\item On the theoretical side, it would be 
very helpful to have estimates
of the amplitudes of the modes.
If relative amplitudes could be reliably
estimated from theory the autocorrelation might be a useful method
for measuring the separation between radial modes and hence 
constrain the density of \dss\ stars.

\item On the observational side we need to be able to
identify the degrees of at least the main modes.
This can in principle be done by comparison of
phases and amplitude ratios for different narrow-band filters as has been
attempted for other \dss\ stars \citep{moya04,dasz05}.

\end{itemize}
We have too few observations in the $B$ filter to follow the latter suggestion, 
and multi-filter (\eg\ Str\"omgren $by$) observations should be considered 
for similar campaigns in the future. The points mentioned here also apply
when planning ground-based support for the
CoRoT \citep{michel06} and Kepler \citep{basri05} satellite missions. 
These missions will provide photometric time series with long temporal coverage
of several \dss\ stars, but only in a single filter.

On request we would gladly provide the light curves presented here.

\section*{Acknowledgments}

This work was partly supported by the Research Foundation Flanders (FWO).
HB was supported by the Danish Research Agency 
(Forskningsr\aa det for Natur og Univers) and
the Instrument center for Danish Astrophysics (IDA).
HB and DS were supported by the Australian Research Council.
This paper uses observations made from the South African Astronomical Observatory,
Siding Spring Observatory in Australia, 
Mount Laguna Observatory operated by 
the San~Diego State University, the Danish 1.54-m telescope at ESO, La Silla, Chile,
Sobaeksan Optical Astronomy Observatory in Korea, 
and Mt.\ Lemmon Optical Astronomy Observatory in Arizona, USA, 
which was operated remotely by Korea Astronomy and Space Science Institute.


\appendix

\section{Data point weights}

\begin{figure}
 \includegraphics[width=8.8cm]{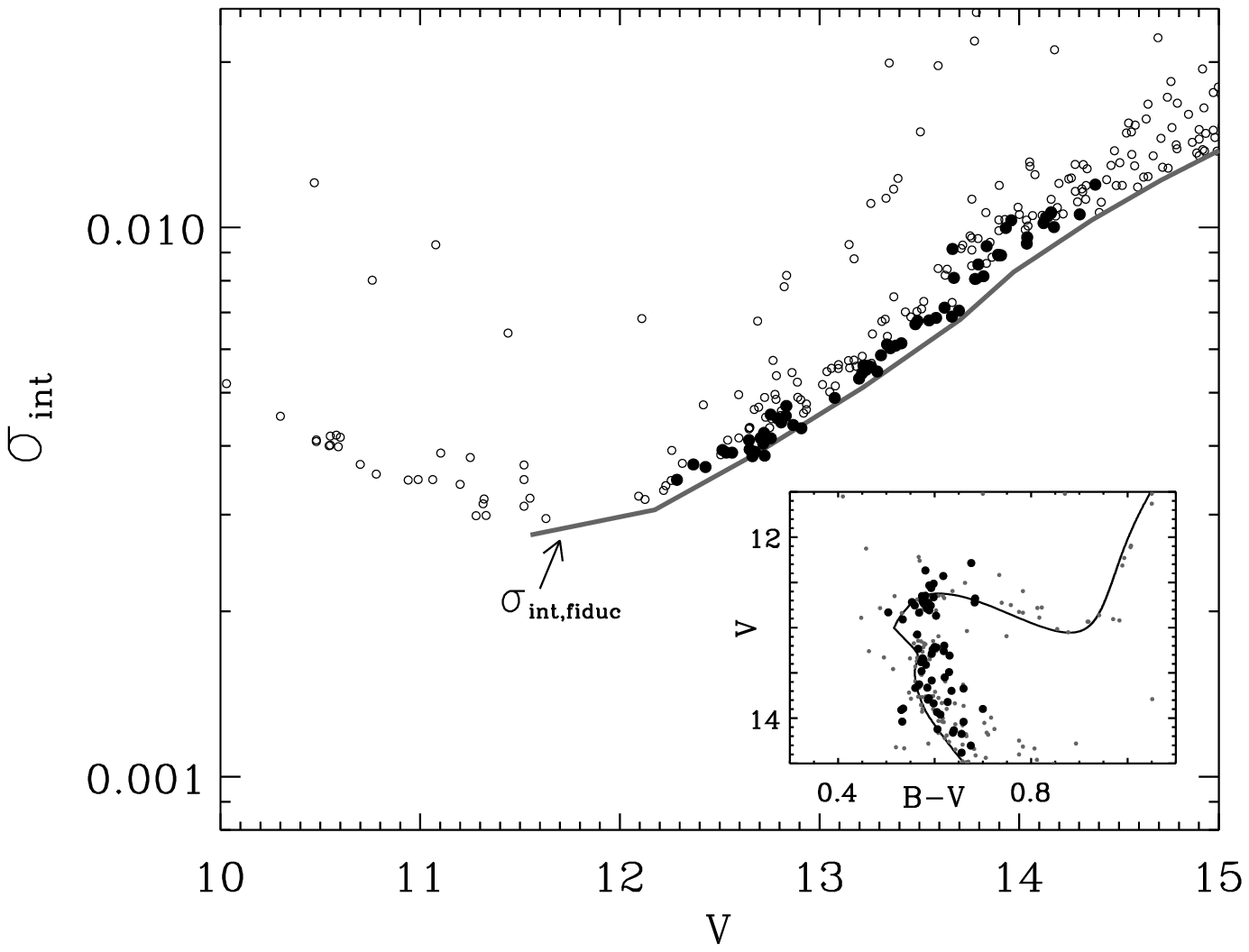}
 \caption{Internal scatter in the light curves 
of bright stars observed at the LOAO site. 
The grey line marks the fiducial $\fid$.
The filled circles mark the reference stars used for measuring weights. 
The inset shows the location of the stars in the \cmd\ of \msyv. 
\label{fig:rms}}
\end{figure}

This Appendix contains a detailed description 
of how we calculated weights using an ensemble of reference stars
(\cf\ Sect.~\ref{sec:offsets}). 
The reference stars were chosen from the most isolated stars, 
\ie\ stars with a maximum of $10$\% flux from neighbouring stars 
within their photometric apertures.
Furthermore, we required that their point-to-point scatter 
was close to the theoretical limit as determined by photon noise. 
The observations from LOAO
had the greatest field-of-view with a total of 358 stars. 
For this site we used $M=65$ reference stars. 
For the site with the smallest CCD we used only 40 reference stars.
In Fig.~\ref{fig:rms} we show the average point-to-point 
scatter (\isd, see Eq.~\ref{eq:isd} below) 
for the stars at the LOAO site plotted \vs\ $V$ magnitude.   
The solid grey line marks the fiducial $\fid$ that represents the 
typical scatter for the best stars at a given magnitude.
The reference stars are marked with filled circles, and
the inset shows the location of the stars in the \cmd.

\subsection{Determination of scatter weights: $\wscat$\label{sec:ptp}}

\begin{figure}
 \includegraphics[width=8.8cm]{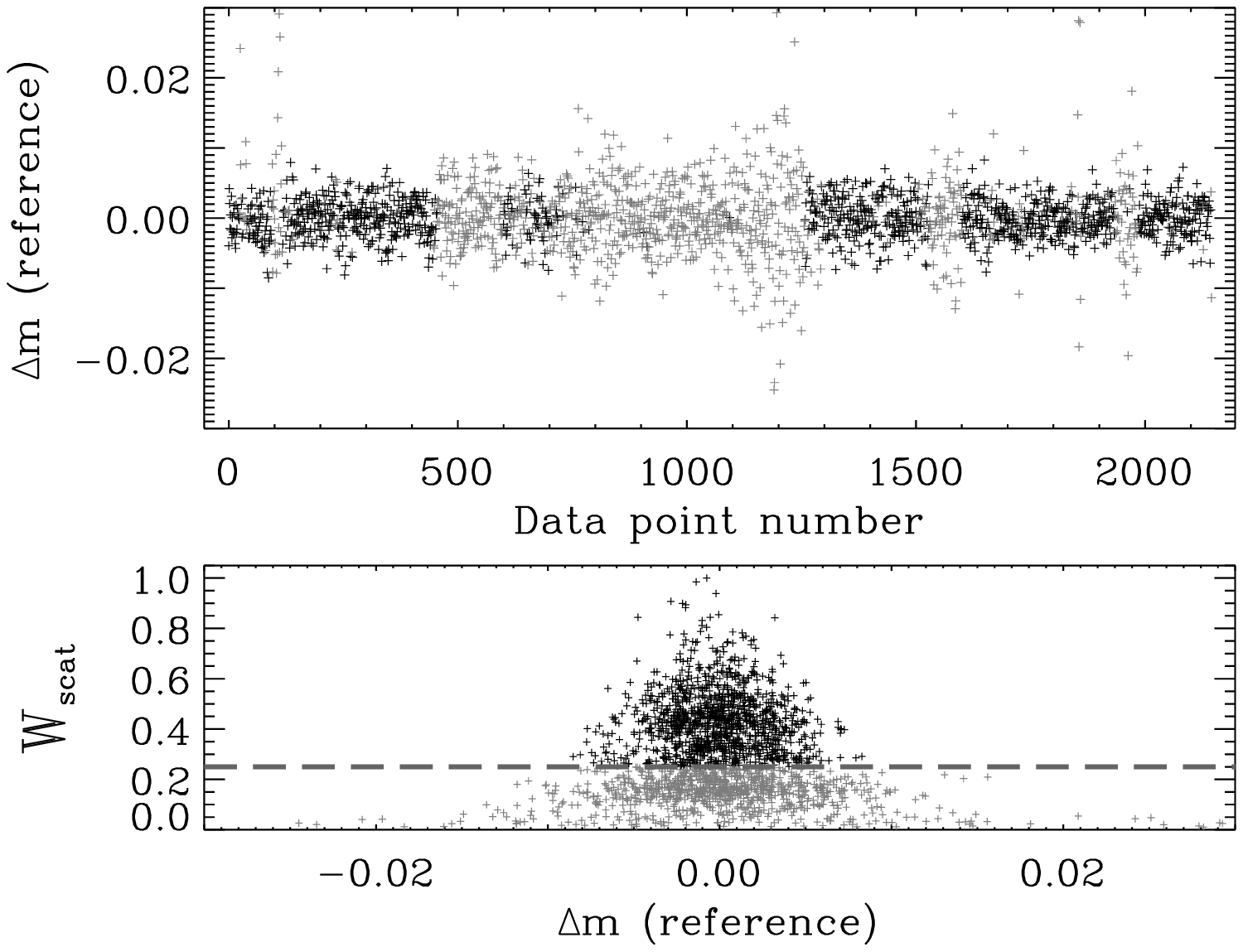}
 \caption{The \topp\ \panel\ shows the light curve of a reference star from WFI.
The grey points have scatter weights $\wscat<25$\%.
The \bott\ \panel\ shows the scatter weight found using all the reference stars. 
The dashed horizontal line marks the $25$\% level.
\label{fig:point}}
\end{figure}


For each reference star we calculated the internal standard deviation 
\begin{equation}
\sigma_{\rm int}^2 = \sum_{k=1}^{N} (m_k - m_{k+1})^2/[2(N-1)], 
\label{eq:isd}
\end{equation}
where $m_k$ is the magnitude of the $k$th measurement and $N$ is the number of data points.
For each reference star \refstar\ and for each data point $k$, we measured 
how much the data point itself and the two 
neighbouring data points ($k-1$ and $k+1$) deviated from zero, as\footnote{The neighbouring data points 
were required to be within a certain time limit, set 
arbitrarily to ten times the median time step between frames (for a given telescope).
If there was only one neighbouring data point, we used only this to
compute the deviation $\delta_\refid$. If there were no
neighbouring data points within the time limit (rarely the case), we assigned a 
weight that was the mean of the weights of data points just before and after
the point in question.}
\begin{equation}
\delta_\refid = {{ |\Delta m_{k+1}| +  |\Delta m_k| + |\Delta m_{k-1}|}  \over {3 \, \fid}},
\label{eq:delt}
\end{equation}
where $\Delta m$ is the deviation from zero and $\fid$ is the fiducial 
scatter (\cf\ solid curve Fig.~\ref{fig:rms}).
Thus, $\delta_\refid$ is a measure of the {\em actual} scatter 
normalized by the {\em expected} scatter for a star of that magnitude.

Based on all $M$ reference stars, we 
then computed the average deviation of the $k$th data point 
as $\langle{\delta}\rangle = {\sum}_{\refid=1}^{M} \delta_\refid / M$,
and calculated the scatter weight as
\begin{equation}
\wscat = {{1} \over ({\langle{\delta}\rangle + \delta_{\rm floor})^n}},
\label{eq:ptpwei}
\end{equation}
where we used $\delta_{\rm floor}=1$ mmag to avoid that a few very good stars dominate.
The optimal choice of the exponent $n$ is discussed in Sect.~\ref{sec:finalwei}.

In Fig.~\ref{fig:point} we show the light curve \vs\ data point number
for one reference star in the \topp\ \panel.
These data are from the Wide-Field Imager (WFI) at Siding Spring, Australia.
In the \bott\ \panel\ we plot the derived scatter weights normalized
so that the highest weight equals 1.        
The dashed horizontal line marks the limit for data points with
scatter weights $\wscat<25\%$. 
These points are plotted with grey symbols in the \topp\ \panel\ as well.

 \begin{figure}
 \includegraphics[width=8.8cm]{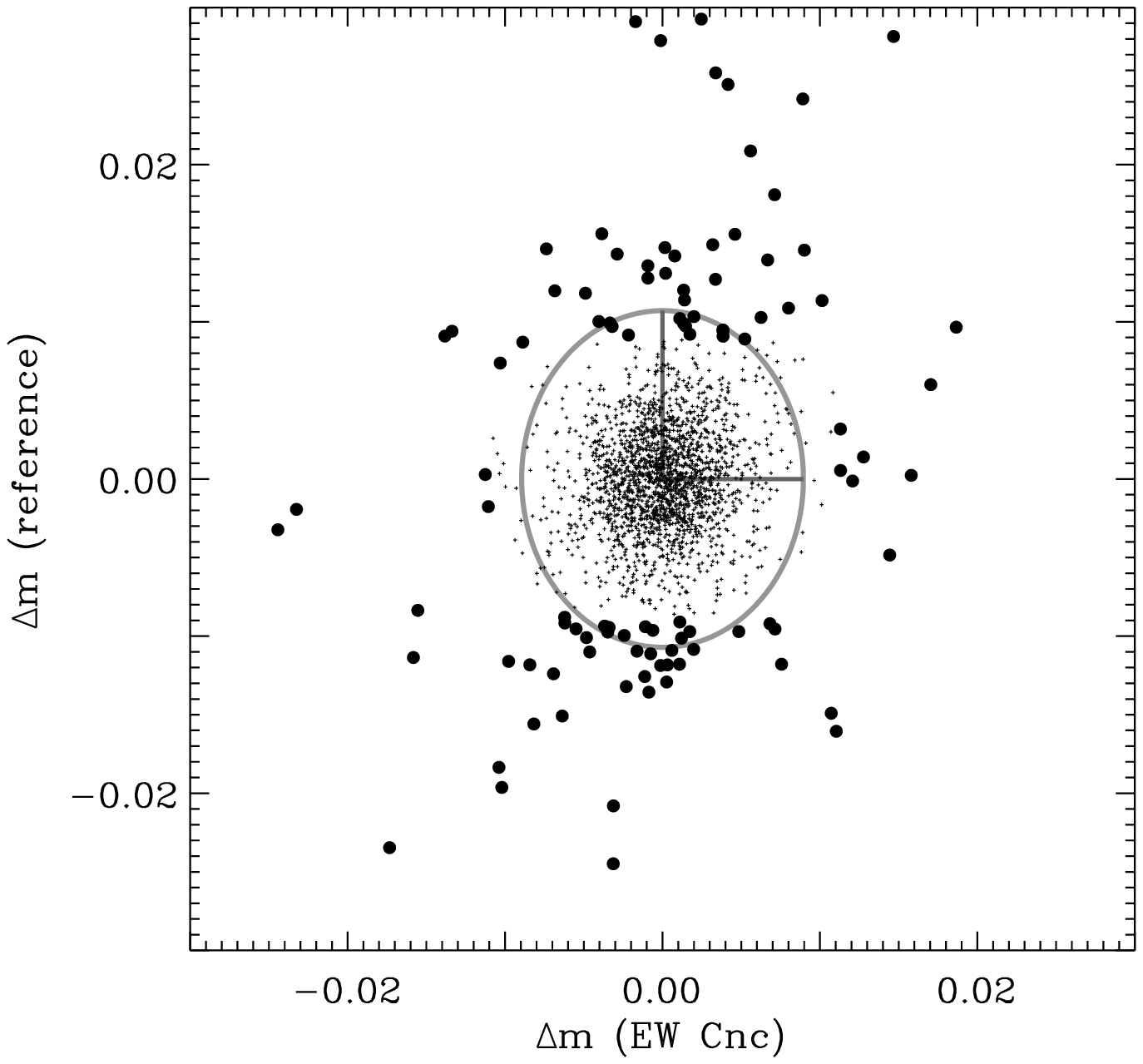}
         \caption{The light curve of a reference star is plotted \vs\ 
                  a target star (EW~Cnc) for data from WFI.  
                  The large black dots have outlier weights $W_\refid<0.5$.
                  We plot the ``3\,$\sigma$ ellipse'' in grey colour (see text for details).
                  \label{fig:outlier1}}
 \end{figure}

\subsection{Determination of outlier weights: $W_{{\rm out}}$\label{sec:out}}

We want to assign low weight to
data points that deviate significantly from the mean magnitude.
We calculated the outlier weight for each data point for the \refstarth\ reference star as
\begin{eqnarray}
W_\refid = \biggl\{ \,1 &+&   \biggl|{{\Delta m_t} \over {a_t \cdot \sigma_t}}\biggr|^b  + \biggl|{{\Delta m_\refid} \over {a_r \cdot \sigma_\refid}}\biggr|^b  \, \biggr\}^{-1},
\label{eq:outlier}
\end{eqnarray}
and the {\em final outlier weight} for the target star 
is the average value $W_{\rm out}=\sum_{\refid=1}^{M}W_\refid/M$ using all $M$ reference stars.
In Eq.~\ref{eq:outlier}, $\Delta m_t$ is the deviation from zero for
the target star and $\sigma_t$ is an {\em estimated} scatter.
Likewise, $\Delta m_\refid$ and $\sigma_\refid$ are the corresponding 
values for the \refstarth\ reference star. 
The values of $a_t$ and $a_r$ determine
how much a data point must deviate before the outlier weight 
becomes significantly smaller than 1.
For the target star we used $a_t=3$ and for the reference stars we used $a_r=2$.
Thus, data points that lie within $3\,\sigma$ for the target star, 
and within $2\,\sigma$ for the majority of the reference stars will
not get low outlier weights. 
The exponent $b$ describes how steeply $W_\refid$
decreases from one to zero. We used the value $b=4$.
To estimate the scatter $\sigma$ in Eq.~\ref{eq:outlier} 
for the target and reference stars, 
we use the scatter weights $\wscat$ from Sect.~\ref{sec:ptp}.
We approximate $\sigma = \sigma_{\rm int} \cdot \eta$, where 
$\sigma_{\rm int}$ is computed from Eq.~\ref{eq:isd} and
$\eta = \langle \wscat\rangle / \wscat$,
where $\langle \wscat\rangle$ is the average scatter weight over the entire series.
Thus, $\eta$ is a measure of the {\em relative} change in scatter from frame to frame.

To illustrate the process we have plotted the light curve of a reference 
star \vs\ the target star EW~Cnc\footnote{Note that the stellar 
oscillations were subtracted before calculating outlier weights.} in Fig.~\ref{fig:outlier1}.
The grey ellipse has half-major and half-semi-major axes corresponding 
to $3\,\sigma_{\rm int}$ for each star.
Intuitively, there are too many points outside this ``3\,\sig\ ellipse''.
If the noise were Gaussian, the fraction of points
that would fall outside the 3\sig\ limit 
would be only $P_{>3\sigma} (\rm 2D) = 1.1\%$.
This corresponds to only 24 out of 2150 data points from WFI, while about 100 are found in the data.
The large black dots in Figure~\ref{fig:outlier1} have
outlier weights $W_\refid<0.5$ and small points have $1\geq W_\refid\geq 0.5$.

 \begin{figure}
 \includegraphics[width=8.8cm]{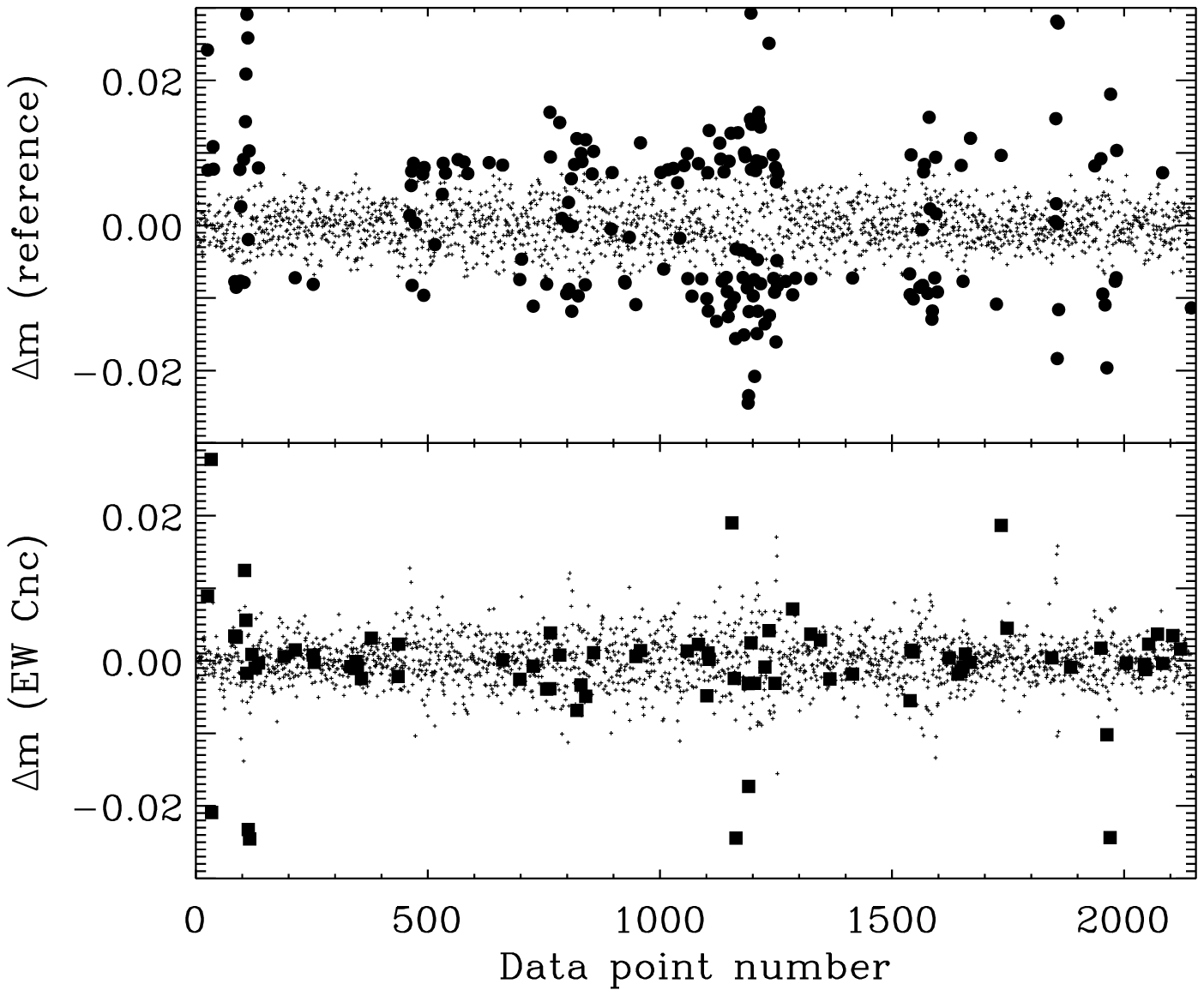}
         \caption{In the \topp\ \panel\ the light curve of a reference star is shown,
                  where filled circles have reference star outlier weights $W_\refid<0.5$.
                  The light curve of the target star (EW~Cnc) is shown in the \bott\ \panel, where
                  filled boxes have final outlier weights $W_{\rm out}<0.5$ (see text for details).
             \label{fig:outlier3}}
 \end{figure}

In Fig.~\ref{fig:outlier3} we show the light curves of EW~Cnc and
the same reference star as in Fig.~\ref{fig:outlier1}.
For the reference star (\topp\ \panel) 
the filled circles mark data points with low outlier weights, \ie\ $W_\refid<0.5$.
The data points of the target EW~Cnc 
are shown in the \bott\ \panel\ in Fig.~\ref{fig:outlier3}. 
We have marked the points with final 
outlier weights $W_{{\rm out}}<0.5$ with filled squares\footnote{Recall that $W_{{\rm out}}$ is the 
mean of $W_\refid$ for all reference stars \refstar.}.
It is interesting to note that several data points are not obvious outliers.
In fact some of the marked outliers have \eg\ $\Delta m<2\sigma$, 
where $\sigma$ is the local {\em rms} scatter. 
The reason for these points having $W_{{\rm out}}<0.5$ is 
that they were outliers in most of the reference stars.

\section{Light curve simulations}\label{sec:vogtere}

Simulating the light curves has two goals:
\begin{itemize}
\item To measure our ability to extract the inserted modes for
cases~A and B and decide which one is optimal.
\item To measure the uncertainty on the frequency, amplitude, and phase.
\end{itemize}

The simulated light curves were constructed to have the same
noise characteristics as for the observations. 
To measure the properties of the noise in the observations,
we first subtracted the detected frequencies before
computing the amplitude spectrum.
We used two components to describe 
the white noise and the drift noise:

\begin{description}

\item[(1)] The white noise was measured
as the mean level in the observed amplitude spectrum
the frequency range 1000--1500\,\mhz.

\item[(2)] The drift noise leads to an increase in the noise level 
towards low frequencies. This ``$1/f$'' noise was measured by fitting 
a fiducial spline function to the observed increase 
in noise in the amplitude spectrum. 

\end{description}

Since each site had a different white noise component and a
different shape of the $1/f$ noise we constructed the light curve
site by site before adding the data as we did for the actual observations.

In practice we created an evenly sampled light curve covering the time span of
the observed time series, but with an over-sampled grid in time by a factor five.
Each light curve consist of random numbers with different seed number. 
The mean is zero and the numbers have a point-to-point scatter 
corresponding to the noise level found in step (1). 
Firstly, we take the Fourier transform of the light curve to the frequency domain,
and multiply the result by the fiducial spline fit found in step (2). 
We then take the Fourier transform to return to the time domain, 
and pick the synthetic data points closest in time to the observed data points. 
Finally, we add the frequencies detected in the star. 
For each star $50$ simulations were realized.

\begin{figure}
\includegraphics[width=8.8cm]{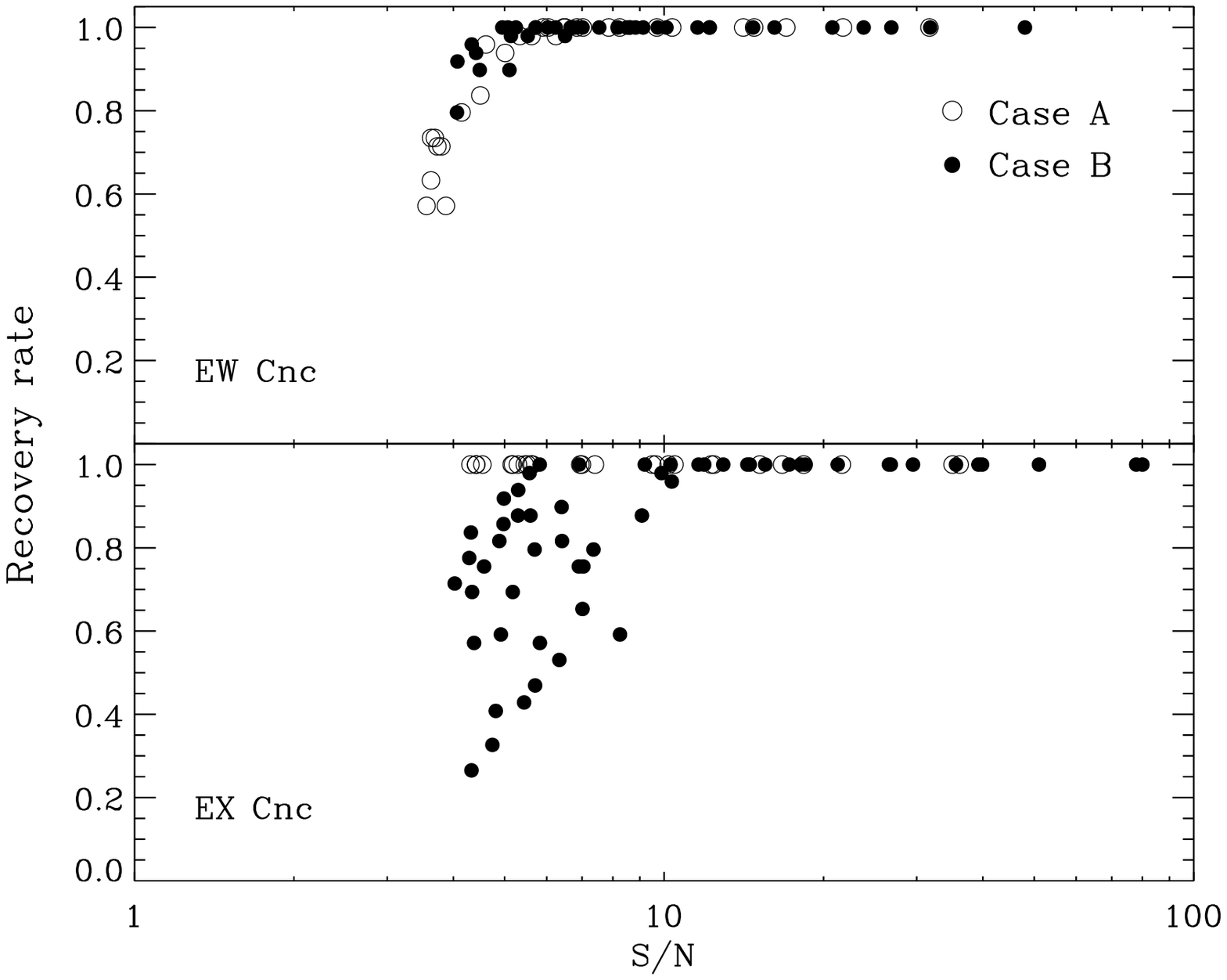}
\caption{Recovery rates of frequencies inserted in the simulations of EW~Cnc (\topp\ \panel)
and EX~Cnc. The results for case~A and B are shown with open and solid symbols, respectively. 
\label{fig:recover}}
\end{figure}

We made a Fourier analysis of each simulation as we did for the observed data (see Sect.~\ref{sec:finalwei}).
In Fig.~\ref{fig:recover} we show how often each inserted frequency 
was recovered in the simulations versus S/N.
The \topp\ and \bott\ \panels\ are 
for EW~Cnc and EX~Cnc, respectively. 
In each panel open and solid symbols correspond to cases A and B (Sect.~\ref{sec:finalwei}).
For EW~Cnc it is seen that in both case~A and B all frequencies with
S/N\,$>5$ are recovered. 
We prefer case~B for EW~Cnc since it has a significantly 
lower residual noise level (see Table~\ref{tab:noise}).
For EX~Cnc we chose case~A because it
shows a much better recovery rate than case~B.

\begin{figure*}
\includegraphics[width=8.8cm]{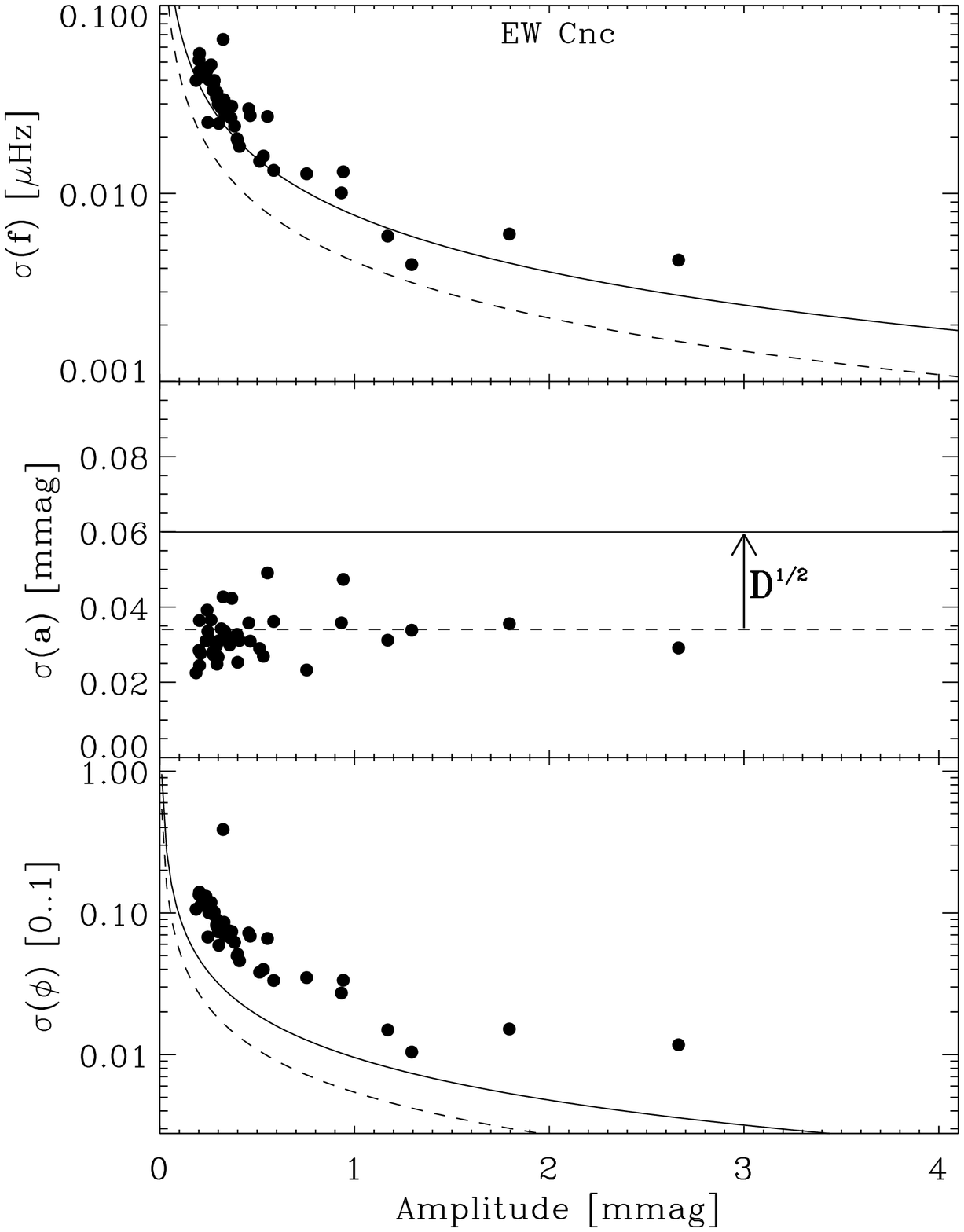}
\includegraphics[width=8.8cm]{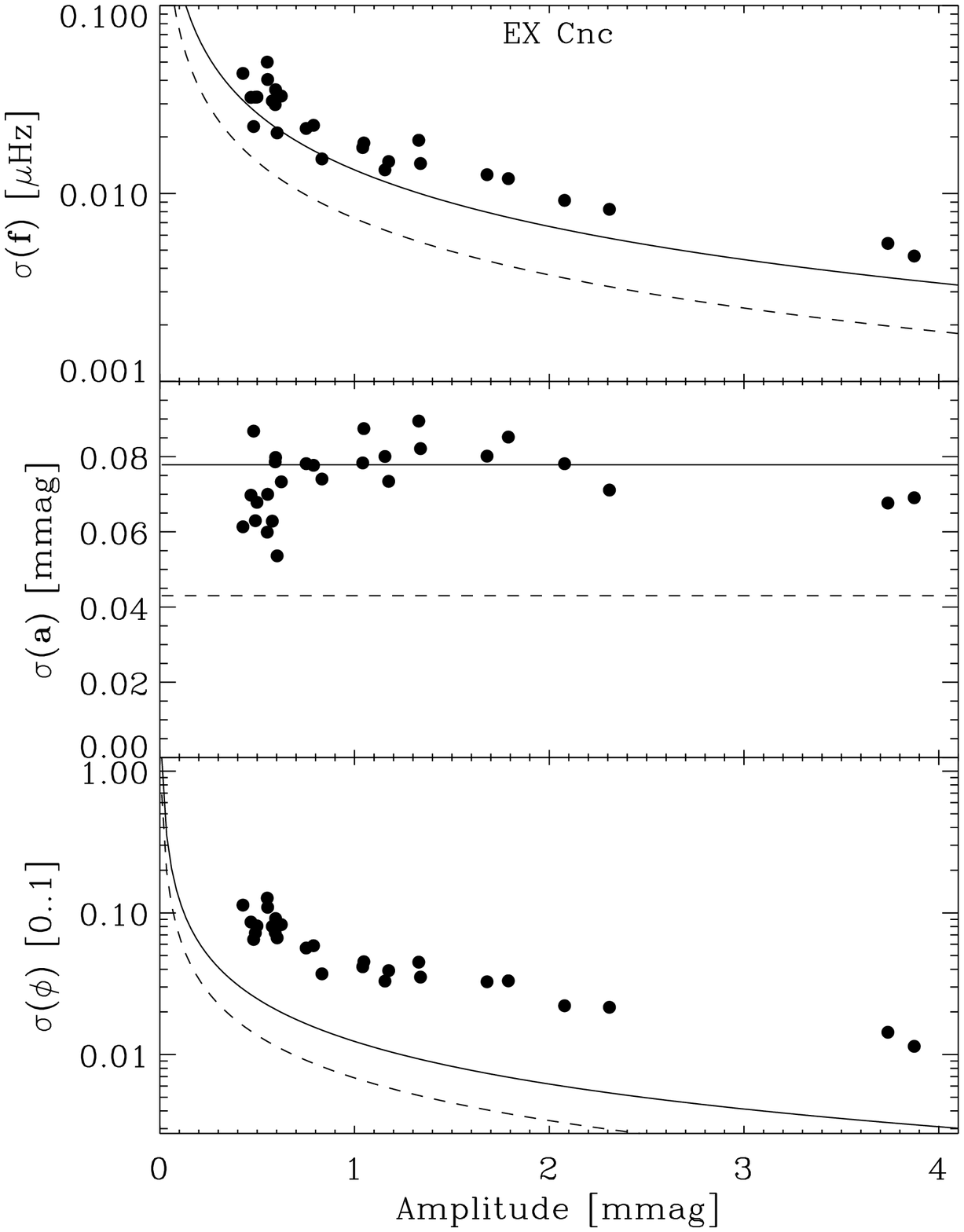}
\caption{Uncertainties of frequency, amplitude, and phase based on simulations 
of EW~Cnc (\lee \panels) and EX~Cnc (\rii \panels).
The dashed lines are theoretical lower estimates from Montgomery \& O'Donoghue (1999), while
the solid lines result from multiplying by the correlation length $\sqrt{D}$.
\label{fig:uncertain}}
\end{figure*}

The uncertainties on frequency, amplitude, and phase given
in Tables~\ref{tab:s1280} and \ref{tab:s1284} 
are computed from the {\em rms} scatter in the simulations.
We plot the uncertainties \vs\ the amplitude for 
EW~Cnc (case~B) and EX~Cnc (case~A) in Fig.~\ref{fig:uncertain}.
The dashed lines are theoretical lower estimates 
of the uncertainties using the expressions from \cite{mont99} which
are valid for uncorrelated noise, 
\ie\ for data with only a white noise component. 
For real data one must multiply the uncertainty 
by the square root of the estimated correlation length \citep{mont99}.
We found this by calculating the autocorrelation of the residual
amplitude spectrum after the mean value has been subtracted. 
The intersection of the autocorrelation function
with zero is a measure of the correlation length, $D$. 
For both EW~Cnc and EX~Cnc we find $D=3.2\pm0.1$ or $\sqrt{D}= 1.79\pm0.06$.
The resulting estimates of the uncertainties 
are marked by the solid lines in Fig.~\ref{fig:uncertain}.

In general we find an acceptable agreement with 
the estimated uncertainties from \cite{mont99}.
For most parameters the estimated uncertainties 
are too low, especially so for the frequencies and phases in EX~Cnc. 
This is likely because the frequencies lie closer in EX~Cnc compared to EW~Cnc.
The uncertainties on the amplitudes in EW~Cnc 
are close to the estimate for uncorrelated noise.

\bsp

\label{lastpage}

\end{document}